# Thermally-driven Neutron Star Glitches

Bennett Link

Department of Physics, Montana State University, Bozeman, Montana 59717;
blink@dante.oscs.montana.edu; and Los Alamos National Laboratory,
MS D436, Los Alamos, NM 87545

and

Richard I. Epstein

Los Alamos National Laboratory, MS D436, Los Alamos, NM 87545;
epstein@sstcx1.lanl.gov

## ABSTRACT

We examine the thermal and dynamical response of a neutron star to a sudden perturbation of the inner crust temperature. During the star's evolution, starquakes and other processes may deposit $\gtrsim 10^{42}$ ergs, causing significant internal heating and increased frictional coupling between the crust and the more rapidly rotating neutron superfluid the star is expected to contain. Through numerical simulation we study the propagation of the thermal wave created by the energy deposition, the induced motion of the interior superfluid, and the resulting spin evolution of the crust. We find that energy depositions of $\sim 10^{40}$ ergs produce gradual spin-ups above the timing noise level, while larger energy depositions produce sudden spin jumps resembling pulsar glitches. For a star with a temperature in the observed range of the Vela pulsar, an energy deposition of $\sim 10^{42}$ ergs produces a large spin-up taking place over minutes, similar to the Vela "Christmas" glitch. Comparable energy deposition in a younger and hotter "Crab-like" star produces a smaller spin-up taking place over $\sim 1$ day, similar to that seen during the partially time-resolved Crab glitch of 1989.

*Subject headings:* stars: evolution — stars: interiors — stars: neutron — dense matter

## 1. INTRODUCTION

Glitches have now been observed in at least 20 pulsars, and are believed to be a phenomenon that every neutron star produces during some period in its evolution. Typical glitches in mature pulsars, *e. g.*, the Vela pulsar, involve fractional jumps in the rotation rate of $\sim 10^{-6}$. Glitches in the relatively young Crab pulsar, on the other hand, are typically a factor of 10 to 100 smaller. The Crab glitch of 1989 was the first whose spin-up was partially time-resolved; following a jump in the spin rate (occurring in < 2 hours), the pulsar completed the remainder of the spin-up over $\sim 1$ d (Lyne, Smith, & Pritchard 1992; see Fig. 6). By contrast, the Vela pulsar has exhibited very different behavior in at least one case. The giant "Christmas glitch" of December 24, 1989 ($1.8 \times 10^{-6}$ fractional spin-up), which occurred during an observing session, could not be time-resolved; most, or possibly all, of the spin-up took place in less than two minutes (McCulloch *et al.* 1990; see Fig. 12). The Crab is unique among the known glitching pulsars in the smallness of its glitches. Giant glitches appear to be typical behavior among mature pulsars. No pulsar has exhibited any unusual behavior just prior to a glitch; these events have always occurred without warning.

Several models for the origin of the glitch phenomenon have been proposed (see, *e. g.*, Ruderman 1969; Baym & Pines 1971; Pines & Shaham 1972; Pines, Shaham, & Ruderman 1972; Anderson & Itoh 1975; Ruderman 1976; Alpar 1977; Greenstein 1979a,b; Anderson *et al.* 1982; Alpar, Cheng, & Pines 1989; Alpar *et al.* 1994). In the starquake model of Ruderman (1969), settling of the crust under gravitational stresses abruptly decreases the star's moment of inertia, increasing the spin rate. At first the starquake model appeared promising, though it soon became clear that starquakes cannot account for the large, frequent (about every 1000 days) glitches of the Vela pulsar (Baym & Pines 1971). Furthermore, since starquakes would occur over sound travel time scales ($\lesssim 0.1$ s), the starquake model is unable by itself to account for the slow 1989 spin-up of the Crab. A more plausible explanation is that these phenomena arise from sudden angular momentum transfer between the crust, whose spin rate we observe, and the more rapidly rotating neutron superfluid of the interior (Anderson & Itoh 1975; Ruderman 1976). Greenstein (1979a,b) originally suggested that sudden heating by a starquake or other source could trigger increased frictional coupling between the crust and the core superfluid. Following the temperature increase, the velocity difference between the superfluid and the crust decreases; the superfluid spins down, while the crust spins up. However, subsequent analyses of timing noise in accretion-powered pulsars (Boynton & Deeter 1979; Boynton 1981; Boynton *et al.* 1984) indicated that $\gtrsim 14\%$ of the neutron star interior responds to torque variations as a rigid body over time scales of 1 to 30 d, requiring at least part of the core to be strongly coupled to the crust. In support of this constraint, theoretical study by Alpar, Langer, & Sauls (1984) suggests that the entire core is tightly coupled to the charged components of the star. In light of these developments, glitches are now believed to originate in the inner crust, where interactions between superfluid vorticity and nuclei create a metastable state in which the superfluid is weakly coupled to the more slowly rotating crust.

Here we consider the problem of the thermal and dynamical response of a neutron star to a sudden *local* increase of the inner crust temperature. We employ the results of recent studies of the frictional coupling between the inner crust superfluid and the crust, which indicate that the coupling is a strongly increasing function of temperature (see, *e. g.*, Alpar 1977; Alpar *et al.* 1984; Alpar, Cheng & Pines 1989; Link & Epstein 1991, hereafter LE; Link, Epstein, & Baym 1993, hereafter LEB; Chau & Cheng 1993a,b). We find that energy depositions of $\sim 10^{40}$ ergs produce gradual spin-ups above the timing noise level, while larger energy depositions produce sudden spin jumps resembling pulsar glitches. For a star with a temperature in the observed range of the Vela pulsar, an energy deposition of $\sim 10^{42}$ ergs produces a large spin-up taking place over minutes, similar to the Vela "Christmas" glitch. Comparable energy deposition in a younger and hotter "Crab-like" star produces a smaller spin-up taking place over $\sim 1$ day, similar to that seen during the partially time-resolved Crab glitch of 1989.

The paper is organized as follows. In §2 we summarize the current understanding of the neutron star interior and give an overview of the problem we are considering. In §3 we discuss the physical conditions of the neutron star inner crust. In §4 we present a descriptions of the superfluid, crust and thermal dynamics. In §5 we describe our numerical simulations of the spin-up process. In §6 we discuss our results, and in §7 we present our conclusions.

## 2. OVERVIEW

### 2.1. The Physical Setting

A neutron star consists of $\sim 1.4 M_\odot$ of mostly neutrons in $\beta$-equilibrium. Beneath a solid or liquid surface, begins the outer crust which contains a lattice of nuclei and relativistically degenerate electrons. At a density of $\sim 4 \times 10^{11}$ g cm$^{-3}$ the inner crust begins, consisting of an $^1S_0$ neutron superfluid, a lattice of neutron-rich nuclei and a sea of relativistically degenerate electrons. Near the density of nuclear matter, $\rho_{nm} = 2.8 \times 10^{14}$ g cm$^{-3}$, the inner crust dissolves into a fluid core consisting of mainly $^3P_2$ superfluid neutrons and a small admixture of superconducting protons with normal electrons. Depending on the the behavior of matter above nuclear density, and the mass of the star, there may exist a distinct inner core containing condensed pions, quarks or some other exotic state.

The neutron superfluid is expected to play a crucial role in the star's rotational and thermal evolution. As a magnetized spinning neutron star slowly loses angular momentum, the external torque acts directly on the charged components, which include the crustal lattice, electrons and protons. The inner crust superfluid, however, couples only weakly to the charged components and maintains a higher rotation rate than that of the crust. Angular momentum transfer from the superfluid to the crust, whether sudden or gradual, affects the star's spin dynamics. Heat generated by the differential rotation between the superfluid and normal components affects the star's thermal evolution (Greenstein 1975; Harding, Guyer, & Greenstein 1978; Alpar *et al.* 1987; Shibazaki & Lamb 1989; Van Riper 1991; Van Riper, Epstein, & Miller 1991; Umeda *et al.* 1993; Van Riper, Link, & Epstein 1995). Coupling between the long-term thermal evolution and the rotational dynamics has been considered by Shibazaki & Lamb (1994) and Shibazaki & Mochizuki (1994).

The rotating superfluid is threaded by an array of vortex lines, aligned with the star's rotation axis. The vortex arrangement determines the rotational state of the superfluid. Within each vortex line core, the fluid is normal. Processes by which a vortex interacts with the charged components include the scattering of electrons, phonons, impurities and lattice defects with thermally excited neutrons in the vortex core (Feibelman 1971; Harding, Guyer, & Greenstein 1978; Jones 1990b), scattering of electrons with the charge distribution induced about the vortex core by its interactions with nuclei (Bildsten & Epstein 1989), coupling of vortex translational motion with the electron-phonon system of the lattice (Jones 1990a), and the coupling of vortex Kelvin modes (kelvons) with lattice phonons (Epstein & Baym 1992; Jones 1992). For angular momentum exchange between the superfluid and the crust to occur through any of these processes, the vortices must change their arrangement. In the inner crust, however, it is expected that interactions between the $^1S_0$ superfluid vortices and the lattice nuclei *pin* the vortices to the lattice, nearly fixing the superfluid angular velocity and allowing a significant velocity lag to develop between the superfluid and the crust as the star slows down (Anderson & Itoh 1975; Alpar 1977; Alpar *et al.* 1984; Epstein & Baym 1988). The superfluid flowing past a pinned vortex lines exerts a radially outward force on it. In the presence of thermal fluctuations, a pinned vortex line occasionally unpins and *creeps* outward (see, *e. g.*, Alpar *et al.* 1984; LE; LEB; Chau & Cheng 1993a,b), the superfluid loses angular momentum and exerts torque on the crust. The coupling between the superfluid and the crust through thermally-activated vortex creep is highly temperature dependent. If the coupling suddenly increases, from a jump in the temperature, for instance, angular momentum transfer to the crust could drive a spin-up.[1]

In contrast to the inner crust superfluid, the core superfluid may be tightly coupled to the charged components. The vortices of the core are strongly magnetized due to entrainment of the superfluid protons by the neutron circulation associated with each vortex (Alpar, Langer & Sauls 1984). Electrons efficiently scatter off the vortex line magnetic moments, coupling the charged components and superfluid together, effectively creating a single fluid. The fluid core is likely coupled to the crust by magnetic stresses (Abney, Epstein, & Olinto 1995).

## 2.2. The Thermally-driven Spin-up

We conjecture that glitches originate from the *sudden* deposition of thermal energy in the inner crust. Because the frictional coupling between the crust and the inner crust superfluid is highly temperature dependent, *any* process that leads to inner crust heating will cause a substantial increase in the coupling. As the star evolves, internal heating can occur through a number of processes, including, internal friction, accretion, plastic or sudden relaxation of structure, and nuclear reactions. Starquakes are expected to occur as the star slows, the centrifugal force at the surface diminishes, and the rigid crust settles. In addition, as the star slows, magnetic flux may be swept from the core by the outward migration of vortex lines. Even though the classical ohmic dissipation time scale is $> 10^{10}$ years for magnetic fields threading the core of a neutron star (see, *e. g.*, Urpin & Van Riper 1993), flux tubes may be expelled much more rapidly. Magnetic pressure in the flux tubes makes them lighter, causing them to float upward (see, *e. g.*, Jones 1987). Another possibility is that the flux tubes are dragged out by the radial motion of neutron-superfluid vortex lines as the star slows. (Srinivasan *et al.* 1990; Ding, Cheng & Chau 1993). In either case the magnetic field topology may evolve in such a way as to produce an abrupt change in direction and strength near the core-crust interface. The magnetic stresses resulting in the crust could induce quakes.

---

[1] Jones (1990c; 1991) has considered the possibility that vortex lines do not pin, and that the superfluid maintains a state of near corotation with the crust. In such a case the superfluid could never acquire a sufficient velocity difference from that of the crust to drive glitches.

For definiteness, we envision the heating occurring suddenly during a starquake. We estimate the *average* accumulation rate of the strain energy from gravitational stresses as (Baym & Pines 1971; Cheng *et al.* 1992)

$$\frac{dE_{\text{strain}}}{dt} \simeq B\theta_c t_{\text{age}}^{-1},  \tag{1}$$

where $B$ ($\sim 10^{48}$ ergs) is the crust strain energy, $\theta_c$ is the critical strain angle at which the lattice breaks, and $t_{\text{age}}$ is the pulsar age. The value of $\theta_c$ is uncertain, and estimates range between $10^{-4}$ (characteristic of the Earth's crust) to 0.1 for a perfect Coulomb lattice. If the accumulated strain energy is released every few years, the Crab and Vela could experience energy depositions in excess of $10^{42}(\theta_c/10^{-2})$ ergs.

Following the energy deposition, a thermal wave propagates through the inner crust, the superfluid of the ever-growing heated region transfers angular momentum to the crust, and the crust spins up. The glitch ends when the first of two things happens: either the thermal energy becomes highly diluted over a large volume, or the superfluid of the active region nearly achieves corotation with the crust. The magnitude and time scale of the spin-up are determined by the relative temperature change caused by the energy deposition. The larger the heating, the greater the coupling rate and the faster the spin-up. Also, the greater the energy deposition, the larger the region that is ultimately affected, and the larger the resulting spin-up.

In the context of this physical picture, we make two remarks:

• *Spin jumps begin suddenly.* The sudden heating that accompanies a starquake causes the spin-up to begin without warning.

• *For a given energy deposition, cooler pulsars produce larger, faster glitches.* As a neutron star cools, its specific heat drops. In a cooler star, with its relatively low specific heat, a given energy deposition gives rise to a relatively large change in the temperature, giving rise to a greater frictional coupling than in the case of a hotter pulsar; the glitch is correspondingly larger and faster.

The first statement is consistent with all glitches detected so far. The second remark is suggestive of the cause behind the markedly different glitch magnitudes and time scales seen during the 1989 Crab glitch and during the Vela Christmas glitch. Before turning to a quantitative description of the thermally-driven spin-up process, we discuss the physical conditions that prevail in the inner crust.

## 3. PHYSICAL CONDITIONS IN THE INNER CRUST

### 3.1. The Equation of State

The nuclear matter equation of state (EOS) determines the bulk properties of a non-rotating neutron star, including the radius, total moment of inertia, inner crust thickness, and inner crust moment of inertia. Recent calculations based on the best potentials currently available for the nucleon-nucleon interaction and including three-body interactions give EOSs that are moderately stiff to stiff (see, *e. g.*, Wiringa, Fiks, & Fabrocini 1989). Moreover, determination of the minimum mass of the inner crust superfluid needed to account for post-glitch relaxation behavior (Link, Epstein, & Van Riper 1992), considered along with the mass measurement of PSR 1913+16 (Wolszczan 1991), rules out very soft EOSs. For the purposes of determining the effective extent and characteristic density of the inner crust superfluid, we will consider the model of Friedman & Pandharipande (1981; FP), a representative EOS of moderate stiffness.

Fig. 1 shows the inner crust density as a function of radius for a $1.4M_\odot$ neutron star built on the FP EOSs. The inner crust ranges in density from $4 \times 10^{11}$ g cm$^{-3}$ to near nuclear matter density, $\rho_{nm} = 2.8 \times 10^{14}$ g cm$^{-3}$. Recent calculations indicate that the the lattice dissolves at a density $\sim 0.6\rho_{nm}$ (Lorenz, Ravenhall, & Pethick 1993; Pethick, Ravenhall, & Lorenz 1995). Most of the moment of inertia of the crust is in the region of highest density, and it is in this region that dynamic coupling between the pinned superfluid and the crust can give the largest contribution to variations in the spin rate of the crust. About 90% of the inner crust mass is in the region $9.8 < r < 10.0$ km with a mean density of $\bar{\rho} \sim 1.5 \times 10^{14}$ g cm$^{-3}$. We take this average density as characteristic of the inner crust superfluid.

### 3.2. Temperature

Soft x-ray emission has been detected from at least four neutron stars. If $T_s$ is the star's measured effective surface temperature, we can estimate the internal temperature, $T$, using the following expression appropriate for a non-magnetized star (Gudmundsson, Epstein, & Pethick 1982)

$$T_8 = 1.288 \left( \frac{T_{s6}^4}{g_{s14}} \right)^{0.455},$$ (2)

where

$$g_s = \frac{GM}{R^2} e^{-\Lambda},$$ (3)

and

$$e^{-\Lambda} = \left(1 - \frac{R_{sch}}{R}\right)^{-1/2}. \tag{4}$$

Here $T_8 \equiv T/10^8$, $T_{s6} \equiv T_s/10^6$, $g_{s14} \equiv g_s/10^{14}$ is the surface gravity, $M$ and $R$ are the stellar mass and radius, and $R_{sch}$ is the Schwarzschild radius.

The effective surface temperature is related to the luminosity seen by a distant observer through,

$$4\pi R^2 \sigma T_s^4 = e^{-2\Lambda} L_\infty. \tag{5}$$

In some cases where surface emission has not been detected (*e. g.*, the Crab pulsar), upper limits on $L_\infty$ are available, and we can obtain an upper limit on the internal temperature using eqs. [2]–[5]. For the purpose of estimating internal temperatures, we identify $T_s$ with the temperature obtained from a black body fit to the observed portion of the spectrum. The dependence of the atmospheric opacity on photon energy, however, can lead to a difference between the temperature inferred from such a blackbody fit and the true effective temperature (see, *e. g.*, Romani 1987; Miller 1992, 1993; Shibanov *et al.* 1993). This issue has yet to be fully resolved.

Fig. 2 summarizes our knowledge of neutron star internal temperatures. Four detections and one upper limit exist in the age range $10^4 - 10^6$ years. These data suggest that stars in this age group have similar internal temperatures.

## 4. DYNAMICAL DESCRIPTION

We consider a star of total moment of inertia $I$, slowing under the influence of a constant external torque, $I\dot{\Omega}_\infty$. The total angular momentum $J$ is given by the angular momentum of the crust and all components strongly coupled to it, plus the angular momentum of the superfluid $J_s$,

$$J(t) = J_c(t) + J_s(t) = I_c\Omega_c(t) + \int d^3r \; r^2\rho_s\Omega_s(r,t) = J_0 - I|\dot{\Omega}_\infty|t, \tag{6}$$

where $J_0$ is the total angular momentum at $t = 0$. We take $I_c$ to be the moment of inertia of the solid crust plus that of the fluid core. The star is in a rotational steady state if the crust and superfluid are both slowing at the same rate, $\dot{\Omega}_\infty$.

### 4.1. The Superfluid

A rotating superfluid is threaded by an array of vortex lines aligned with the rotation axis, each contributing a quantum of vorticity $\kappa$ to the net superfluid flow. [For a neutron superfluid, $\kappa = h/2m_n$ where $m_n$ is the neutron mass]. For a superfluid with an average flow velocity $\vec{v}_s(r)$ and angular velocity $\Omega_s(r)$, the total circulation is given by the total number of vortex lines within an integration contour enclosing the axis of rotation,

$$\oint \vec{v}_s \cdot d\vec{l} = 2\pi r^2 \Omega_s(r) = \kappa \int_0^r dr' 2\pi r' n(r'),$$ (7)

where $r$ is measured from the rotation axis and $n(r)$ is the local areal density of vortex lines. Differentiating eq. [7] with respect to $r$, we obtain a differential equation relating the vortex density to the local superfluid angular velocity,

$$\kappa n(r) = 2\Omega_s(r) + r \frac{\partial \Omega_s(r)}{\partial r}.$$ (8)

Radial vortex motion changes the number of vortex lines within a radius $r$, and hence the superfluid rotation rate (see eq. 7). For vortex lines crossing a cylindrical boundary at $r$ with a radial velocity component $v(r)$, number conservation requires

$$\frac{\partial N}{\partial t} + n(r) \ 2\pi r v(r) = 0,$$ (9)

where $N$ is the total number of vortex lines contained within the volume. An equivalent expression is

$$\frac{\partial \Omega_s(r, t)}{\partial t} = -v \left( \frac{2}{r} + \frac{\partial}{\partial r} \right) \Omega_s(r, t).$$ (10)

Eq. [10] is the basic dynamical equation describing the superfluid rotation. The radial component of the vortex velocity, $v$, is determined by the manner in which pinned vortices unpin and move through the inhomogeneous environment of the crustal lattice, a problem which has been studied in various limits (LE; LEB; Chau & Cheng 1993a,b). We use here the results of LE and LEB, which have the most general applicability. A pinned vortex segment unpins by overcoming an activation barrier whose height decreases as the lag velocity between the superfluid and the crust, $\omega \equiv \Omega_s - \Omega_c$, is increased. The mechanics of the unpinning process, and hence, the height of the activation barrier, is determined by $\omega$ and by the importance of vortex tension relative to the pinning force per nucleus. Estimates of the vortex stiffness (LEB) based on the gap calculations of Ainsworth *et al.* (1989) indicate that vortices are very stiff in the densest regions of the inner crust, such that unpinning of a vortex segment in these regions requires the breaking of multiple pinning bonds simultaneously. As the lag $\omega$ is reduced, more pinning bonds must break, and the activation energy grows as $\omega^{-1}$. For the neutron star internal temperatures of interest here, stiff vortices overcome their pinning barriers primarily through classical thermal activation, rather than through quantum tunneling. We use the following expression for the thermally-activated vortex creep velocity (eq. 6.9, LEB with eq. B.12, LE):

$$v(\omega, T) = v_0 \ e^{-\beta/T\omega}.$$ (11)

This form applies in the limit $\omega \ll \omega_c$, where $\omega_c$ is the critical lag at which vortices cannot remain pinned; our use of this limit is justified below. The quantity $\beta$ measures the strength of the coupling between the superfluid and the normal matter; it depends only on the local mass density, and is the same for a region of a given density for all stars. The coupling parameter can be written in terms of $\omega_c$ and the effective pinning strength $E_p$ per nucleus, and can also be related to observable quantities (eqs. 6.11 and 6.14, LEB):

$$\beta = 0.54 \ \omega_c E_p \simeq |\dot{\Omega}_\infty| t_r(T_0) T_0 \ln^2\left(\frac{4v_0 \ t_{\mathrm{age}}}{R}\right), \tag{12}$$

where $t_{\mathrm{age}} \equiv \Omega_s/2|\dot{\Omega}_\infty|$. Here $T_0$ is the temperature in the absence of sudden heating and $t_r$ is the coupling time between the superfluid and the normal matter. The dependence of $\beta$ on $t_{\mathrm{age}}$ and $T_0$ in eq. [12] is only apparent; $t_r$ scales as $(|\dot{\Omega}_\infty| T_0 \ln^2[4v_0 t_{\mathrm{age}}/R])^{-1}$ (LEB). For a relaxation time measured from post-glitch spin behavior and an internal temperature deduced from surface thermal emission, $\beta$ can be determined. The exact value of $v_0$ is unimportant, and is expected to be $\sim 10^6$ cm s$^{-1}$ (LE; Epstein & Baym 1992; LEB).

If the inner crust temperature is perturbed above $T_0$, the vortex velocity increases dramatically, the superfluid rotation rate decreases (see eq. 10), and the crust spins up. As the superfluid approaches corotation with the crust, the vortex velocity drops precipitously, and the superfluid is essentially *decoupled* from the crust. Immediately after the glitch, the external torque acts on a lesser moment of inertia, and the crust spins down at a *greater* rate.

## 4.2.  Thermal Dynamics

The time-dependent coupling of the superfluid to the crust is determined by the evolution of the local temperature. Following the sudden energy deposition, a thermal wave moves through the crust according to

$$C_v \frac{\partial T}{\partial t} = \vec{\nabla} \cdot (\kappa_T \vec{\nabla} T), \tag{13}$$

where $C_v$ and $\kappa_T$ are the local specific heat and thermal conductivity, both functions of temperature and density. In eq. [13] we neglect the heat generated by the friction between the two components; as discussed in §6, heat generation is unimportant for stars in the temperature range of the Crab and Vela, but is likely to be important for colder stars. The specific heat resides mostly in the degenerate electrons and scales linearly with temperature. Thermal conduction is inhibited by phonons, and the conductivity decreases with increasing temperature.

## 5.  SIMULATIONS: GEOMETRY AND INITIAL CONDITIONS

We model the portion of the inner crust superfluid that drives the spin-up as a cylindrical shell. The greatest contribution to the moment of inertia of the inner crust superfluid is in a band about the equator of the star, and a cylindrical geometry simplifies our calculations while capturing the essence of the inner crust dynamics. Furthermore, we expect crust cracking in a slowing neutron star to occur mainly near the equator, where the gravitational stresses become the greatest. A full reproduction of the many aspects of pulsar timing behavior is beyond the scope of our simple model. The multi-component nature of post-glitch recovery is presumably due to the contributions of distinct inner crust regions which find themselves in different rotational states following the glitch (see, *e. g.*, Alpar *et al.* 1984). We will focus here instead on reproducing the spin-ups themselves, which occur more rapidly than the post-glitch recovery.

The effective crust in our model is a cylindrical shell of constant superfluid mass density $\bar{\rho}$, thickness $\Delta$, and height $h$, with outer and inner radii $R_c \pm \Delta/2$ (see Fig. 3). We define a cylindrical coordinate $x$, such that $x = 0$ at radius $R_c$. We estimate the height of the effective pinning region by the height at which at a vortex a distance $R_c$ from the axis exits a spherical shell of radius $R_c + \Delta/2$. For an FP EOS, $h$ is $\sim 0.3 R_c$. The height of the heated region is more difficult to estimate, but could be comparable to that of the pinning region, as we assume.

We consider the deposition of energy $E$ in the cylindrical effective crust, such that the greatest heating is at the center of the cylinder, and the temperature falls off towards its inner and outer boundaries. The heating is independent of position along the rotation axis. The temperature distribution is in a gaussian profile of half-width $\sigma$ in the $x$ direction, *i. e.*,

$$\Delta T(x, t = 0) = \Delta T(0,0) \mathrm{e}^{-x^2/2\sigma^2}. \tag{14}$$

For a specific heat that depends linearly on $T$, the energy deposited is given by

$$
\begin{aligned}
E &= 2\pi \int_{-h/2}^{h/2} dz \int_{R_c-\Delta/2}^{R_c+\Delta/2} dr \int_{T_0}^{T} dT \; C_v(T) \\
&\simeq 2\pi R_c h C_{v0} T_0 \int_{-\infty}^{\infty} dx \left\{ \frac{\Delta T(x, t=0)}{T_0} + \frac{1}{2} \left( \frac{\Delta T(x, t=0)}{T_0} \right)^2 \right\},
\end{aligned}
\tag{15}
$$

where the subscript 0 denotes quantities evaluated at the unperturbed temperature, $T_0$. We take $\sigma \ll \Delta$, and have extended the integration over an infinite domain as a simplification. Additionally, we neglect the effects of thermal diffusion into regions of lower density.

The solution to eq. [15] is

$$\frac{\Delta T(0,0)}{T_0} = -\sqrt{2} + \left( 2 + \frac{E}{\pi^{3/2} C_{v0} T_0 R_c h \sigma} \right)^{1/2}. \tag{16}$$

We are interested in situations where $\sigma \ll h < R$, so considering one-dimensional diffusion of a planar wave in the $x$ direction is a good approximation. In this preliminary study we neglect the effects of the temperature dependence of the thermal conductivity and specific heat on the evolution of the diffusion wave, and solve the diffusion eq. [13], which is now

$$\frac{\partial T}{\partial t} = \frac{\kappa_0}{C_{v0}} \frac{\partial^2 T}{\partial x^2}. \tag{17}$$

This linear approximation is rigorously valid only for temperature enhancements significantly less than unity; however, below we show this approximation to be adequate for temperature enhancements of order unity as well. The solution to eq. [17] with our initial conditions is

$$\Delta T(x, t) = \Delta T(0,0) \left( \frac{t_0}{t} \right)^{1/2} e^{-t_0 x^2 / 2t\sigma^2}, \qquad (18)$$

where $t_0 \equiv C_{v0} \sigma^2 / 2\kappa_0$.

Assuming that the superfluid and the crust are in rotational equilibrium before the energy deposition, both components slow at the same rate $\dot{\Omega}_\infty$. To obtain the initial lag throughout the crust, we solve eqs. [10] and [11] numerically with $\dot{\Omega}_s(r, t) = -|\dot{\Omega}_\infty|$. We then deposit an amount of energy $E$, and solve eqs. [16] and [18] for the temperature throughout the crust as a function of time. At each time step, we solve eqs. [10] and [11] for $\Omega_s(r, t)$, eq. [6] for $\Omega_c(t)$, and then update $\omega(r, t)$.

# 6. DISCUSSION

## 6.1. Example Simulations

To illustrate the consequences of sudden inner crust heating, we present solutions for three models (see Tables 1 and 2). For the first model, the parameters were chosen as representative of the Crab pulsar. The second model represents Vela, and the third describes an older star, $5 \times 10^5$ years in age. The Crab's surface temperature is taken to be about half the observational upper limit. The temperatures of Vela and the older star are the same, about one-third that of the Crab. The superfluid coupling parameter depends on the strength of vortex pinning only, and to the extent that spin-up events in different pulsars are driven by regions of the same density, we expect $\beta$ to be a constant among pulsars; we take $\beta = 2 \times 10^8$ rad s$^{-1}$ K in each of our three models. For this choice, the superfluid relaxation times implied by eq. [12] for our assumed temperatures are 495 d for Vela and $\sim 7$ d for the Crab, consistent with the post-glitch relaxation times seen in these two pulsars. The lag velocities in the steady state are 0.04 rad s$^{-1}$ for the simulation of the Crab, and $\lesssim 0.12$ rad s$^{-1}$ for the simulations of Vela and an older pulsar. As these lag velocities are significantly less than the estimates of $\omega_c \simeq 1$ rad s$^{-1}$ in the densest regions of the crust (LEB, Table 1, using the gap parameters of Ainsworth *et al.* 1989), our use of the low-lag form for the vortex velocity (eq. 11) is justified. The stellar radius, moment of inertia, effective crust extent, and average superfluid density are based on the FP EOS. For an FP star, 90% of the crust's mass is contained in the region $9.8 < r < 10.0$ km; we take $\Delta$, the thickness of the effective crust to be 200 m, and $R_c$, the radius of the center of the effective crust, to be 9.9 km. The initial width of the heated region is $\sigma = 20$ m for each model. We now discuss example simulations from each of our three models.

**Crab-like Simulation.** As our first example, we consider an energy deposition of $2.1 \times 10^{42}$ ergs. The progress of the thermal wave during the glitch is shown in Fig. 4. The energy deposition produces a two-fold increase of the initial inner crust temperature. As the thermal wave propagates, the superfluid in the heated region loses angular momentum (see Fig. 5). The angular momentum lost from this region is given to the crust, whose spin behavior we show in Fig. 6. About half of the angular momentum transfer occurs within the first 100 seconds. The remainder of the spin-up occurs over the following day, and amounts to a final spin-up of $\Delta\Omega_c/\Omega_c \simeq 7 \times 10^{-8}$. After $\sim 1$ d, the thermal energy has been diluted through the crust, and the superfluid relaxes over the following days, during which the crust spins down at a *greater* rate than before the glitch. During this time, the superfluid in the region surrounding $x = 0$ has reduced the lag to the degree that vortex motion is effectively turned off, the superfluid of this region is decoupled from the crust, and the external torque acts on a lesser total moment of inertia. The temperature change produces only small changes in the local vortex density, depletions and accumulations of less than 1% (see Fig. 7). In spin-up magnitude and time scales, as well as post-glitch relaxation, this simulated spin-up is similar to the 1989 Crab glitch.

**Vela-like Simulation.** This model differs from the model for the Crab in the unperturbed temperature (about three times lower), and the star's angular velocity and spin-down rate. For this example, the energy deposition is $1.51 \times 10^{42}$ ergs. Because of the lower specific heat of this cooler pulsar, the energy deposition produces a nearly six-fold increase in the initial temperature (Fig. 8), resulting in a significantly larger and faster change in the superfluid rotation rate (Fig. 9). Roughly half of the angular momentum transfer occurs within the first second, with the remainder occurring over the subsequent several hundred seconds, giving a final spin-up of $\Delta\Omega_c/\Omega_c \simeq 2 \times 10^{-6}$ (Fig. 10). Significant vortex motion occurs, with vortex depletions and accumulations of order unity (see Fig. 11). This simulated spin-up is consistent in both magnitude and time scale with the Vela Christmas glitch. In Fig. 12 we show the predicted pulse arrival times, compared to the data from the Christmas glitch. Lesser energy input produces a smaller, slower glitch; an example is shown in Fig. 13. Small events of this short have been seen in Vela.

**An Older Pulsar.** In Fig. 14 we show the response of an older, more slowly decelerating star to an energy deposition of $9 \times 10^{41}$ ergs. For an energy deposition of $1.51 \times 10^{42}$ ergs, as considered in the Vela-like simulation, the spin-up is $\sim 2.5$ times larger, over a shorter time scale.

An interesting question is how the glitch magnitude and time scale depend on the energy deposition; these quantities are shown in Figs. 15–17. Spin-ups above the timing noise level ($\Delta\Omega_c/\Omega_c \sim 10^{-10}$) occur for energy depositions above $\sim 10^{40}$ ergs. For energy depositions $\sim 10^{42}$ ergs and larger, a given energy deposition produces a spin-up nearly two orders of magnitude larger in the Vela-like simulation than in the Crab-like simulation. In the simulation of an older pulsar, the behavior at low energies is similar to that of the Crab. For large energy deposition, however, the resulting spin-up is somewhat larger than in the model of Vela. For very large energy depositions ($\sim 10^{43}$ ergs) the curves in Fig. 15 begin to flatten out; in these cases the energy deposition causes nearly the entire inner crust superfluid to achieve corotation with the crust, giving the largest possible glitch. Energy depositions of such magnitude are significantly greater than those expected from starquakes. In Figs. 16 and 17 we show the fraction of the total spin-up after selected times as a function of energy deposition. The rate of the spin-up increases markedly with energy, and hence, with glitch magnitude.

## 6.2.   General Properties of the Solutions

**Dependence on temperature.** The quantity that most strongly affects the spin-up magnitudes and time scales among our simulations is $C_{v0}T_0 \propto T_0^2$, which determines the relative temperature change for a given energy deposition (see eq. 16); it is a factor of $\sim 10$ lower in our Vela-like and old pulsar simulations than in the Crab-like simulation. Consequently, a given energy deposition produces a larger (up to a factor of 10) temperature change in the Vela-like and older pulsar simulations, leading to a larger and faster transfer of angular momentum from the superfluid to the crust. The glitch is correspondingly larger and faster.

**Dependence on the superfluid coupling parameter.** The glitch magnitude depends sensitively on the magnitude of the superfluid coupling parameter $\beta$. For a given value of $\beta$ the lag in the steady state, prior to the energy deposition, is from eqs. [10] and [11]

$$\omega_{ss} \simeq \frac{\beta}{T_0} \ln^{-1} \frac{4 v_0 t_{\text{age}}}{R}. \tag{19}$$

For larger $\beta$, the superfluid is less tightly coupled to the crust and $\omega_{ss}$ is larger. After the energy deposition, the local vortex velocity becomes

$$v(\omega, T) = v_0 \exp\left[-\frac{\beta}{\omega_{ss}T(1 + \delta\omega/\omega_{ss})}\right], \tag{20}$$

where $\delta\omega$ is the local change in the lag from $\omega_{ss}$ as the spin-up proceeds ($\delta\omega < 0$). For larger $\beta$ and $\omega_{ss}$, the vortex velocity decreases less rapidly as the lag decreases, and the glitch is correspondingly larger.

**Dependence on pulsar age.** An older pulsar has a lower vortex creep velocity in the steady state and a lower $\omega_{ss}$. The vortex velocity in an older star increases less with the temperature increase, and drops more rapidly as the lag decreases (see eq. 20). For small or modest energy input, the spin-up is smaller (compare the Vela-like and old pulsar simulations in Fig. 15). For a pulsar of any age, however, a very large energy deposition brings the entire inner crust superfluid to a state of near corotation with the crust, giving rise to a glitch of magnitude

$$\frac{\Delta\Omega_c}{\Omega_c} \simeq \frac{I_s \omega_{ss}}{I_c \Omega_c}, \tag{21}$$

where $I_s$ is the inner crust moment of inertia. The glitch magnitude in an old pulsar is now larger than in a younger pulsar of the same temperature by a factor of up to $\Omega_c^{-1}$. [The decrease in $\omega_{ss}$ with pulsar age is relatively small]. In our model of an old pulsar, the spin rate is $\sim 7$ times smaller than in our model of Vela, corresponding to a glitch larger by the same factor for energy inputs of $\sim 10^{43}$ ergs (see Fig. 15).

**Dependence on dimensions of initial heating.** The total spin-up and the time scale over which it occurs are rather insensitive to the initial width of the energy deposition, $\sigma$; the glitch magnitude changes by a factor of $\sim 2$ as $\sigma$ is varied from 5 m to 100 m. The glitch magnitude increases slowly with the height of the effective region, $h$.

### 6.3. Other Effects

**Non-linear diffusion.** The large ($\sim 6$-fold) increases in the local temperature we have considered in our Vela-like simulation raise the question of the adequacy of our linear treatment of the diffusion problem. Accounting for the temperature dependence of the heat transport would quantitatively change our results. To estimate the importance of these non-linear effects, we have performed a simulation with our Vela-like model using a thermal diffusivity that is smaller by a factor of $\sim 8$, corresponding to a temperature $\sim 3$ times higher. The initial 300 seconds of the evolution closely follow the results of the original simulation at lower temperature. Owing to the lower thermal diffusivity, the final portion of the spin-up occurs over $\sim 1$ hour. The resulting spin-up is about 25% larger because more vortices can move before this more slowly moving thermal wave decays. We expect the results presented here to be qualitatively unchanged by the real effects of non-linear diffusion.

**Frictional heating.** In this study we have ignored the back-reaction of the frictional heating that accompanies the glitch. This heating further increases the frictional coupling. An important question to be resolved is if a feedback process could lead to a thermal runaway, as suggested by Greenstein (1979a,b), wherein an enormous spin jump occurs, accompanied by a "thermal explosion" from the stellar surface. The effects of glitch-induced heating become important when the heat generated is comparable to the initial energy deposition. For a lag $\omega$ prior to the glitch, the heat liberated is $\Delta J \omega$, where $\Delta J$ is the change in angular momentum. The lag depends on the strength with which vortices pin to the inner crust lattice, and we estimate that it is $\lesssim 10$ rad $s^{-1}$ (LE). In the Vela pulsar $\Delta J$ is of the order of $10^{41}$ erg rad$^{-1}$ s, typically, and the heat generated during the glitch is $\lesssim 10^{42}$ ergs, less than the energy deposition considered in our example simulation. However, we expect that in a cold pulsar, with its correspondingly lower specific heat, giant glitches can arise from energy depositions significantly less than $10^{42}$ ergs; we anticipate the effects of frictional heating to play an important role in the rotational and thermal dynamics of these pulsars. We will consider this possibility in future work.

### 7. CONCLUSIONS

As a neutron star evolves, any process that leads to internal heating will affect both the thermal and rotational evolution of the star. In particular, starquakes could produce significant internal heating, dramatically increasing the frictional coupling between the crust and the more rapidly rotating neutron superfluid of the interior. In this paper we have studied the thermal and dynamical response of a neutron star to a sudden deposition of energy in the inner crust, as would result from a starquake. We find that energy depositions of $\sim 10^{40}$ ergs produce gradual spin-ups above the timing noise level, while larger energy depositions produce sudden spin jumps resembling pulsar glitches. The differences in spin behavior between the Crab pulsar and other, more mature pulsars may be attributable to differences in temperature. For a star with a temperature in the observed range of the Vela pulsar, an energy deposition of $\sim 10^{42}$ ergs produces a large spin-up taking place over minutes, similar to the Vela "Christmas" glitch. Comparable energy deposition in a younger and hotter "Crab-like" pulsar produces a smaller spin-up taking place over $\sim 1$ day, similar to that seen during the partially time-resolved Crab glitch of 1989.

## ACKNOWLEDGEMENTS

We thank K. Van Riper for helpful discussions, and for providing us with the inner crust specific heat and thermal conductivity. This work was performed under the auspices of the U.S. Department of Energy, and supported in part by NASA EPSCoR Grant #291511.

Fig. 1—Inner crust density versus distance from the stellar center for a star constructed from an FP EOS. The dashed line indicates a density of $0.6\rho_{nm}$, at which the calculations of Lorenz, Ravenhall, & Pethick (1993) indicate the nuclei dissolve.

Fig. 2—Neutron star internal temperatures. Note the similarity in temperatures in the age range $10^4 - 10^6$ years. Except for the Crab pulsar, for which the age is known, an $n = 3$ breaking index model was used to obtain the pulsar age. An FP EOS was assumed.

Fig. 3—Model geometry. The effective crust is modeled as a cylindrical shell. Heating occurs in a region of height $h$ and thickness $2\sigma$.

Fig. 4—Evolution of the inner crust temperature enhancement in our Crab-like simulation, following the deposition of $2.1 \times 10^{42}$ ergs.

Fig. 5—Evolution of the superfluid. We show the change in the superfluid angular velocity, relative to its value before the energy deposition.

Fig. 6—The response of the crust following the deposition of $2.1 \times 10^{42}$ ergs at $t = 0$. The pre-glitch spin evolution has been subtracted. The data from the 1989 glitch are shown for comparison. About half of the spin-up occurs during the first two minutes. At late times the superfluid is decoupled from the crust, the external torque acts on a lesser moment of inertia, and the spin excess decays.

Fig. 7—Evolution of the vortex density, in units of the initial density $n_0$.

Fig. 8—Evolution of the inner crust temperature enhancement in our Vela-like simulation, following the deposition of $1.5 \times 10^{42}$ ergs.

Fig. 9—Evolution of the superfluid.

Fig. 10—The response of the crust following the deposition of $1.5 \times 10^{42}$ ergs at $t = 0$. The pre-glitch spin evolution has been subtracted. About half of the spin-up occurs during the first three seconds. Decay of the spin excess, not visible on this scale, occurs at late times.

Fig. 11—Evolution of the vortex density.

Fig. 12—The pulse arrival times for the simulation of Fig. 10. The arrival times expected from the pre-glitch behavior have been subtracted. Data from the Christmas glitch are shown.

Fig. 13—A small, slow spin-up and the subsequent recovery resulting from the deposition of $2 \times 10^{40}$ ergs in our model of the Vela pulsar.

Fig. 14—The response of an older star (age $5 \times 10^5$ years) to the deposition of $9 \times 10^{41}$ ergs at $t = 0$. The unperturbed temperature is equal to that of our Vela-like simulation.

Fig. 15—Total spin-up as a function of energy deposition.

Fig. 16—Fraction of the total spin-up versus glitch magnitude at selected times, for our model

of the Crab pulsar. The upper and lower data points correspond to the total spin-ups observed during the 1989 Crab glitch after 1 d and 0.1 d, divided by the spin-up achieved after 1.64 d (the highest data point appearing in Fig. 6).

Fig. 17—Fraction of the total spin-up versus glitch magnitude at selected times, for our model of the Vela pulsar.

## FP Inner Crust

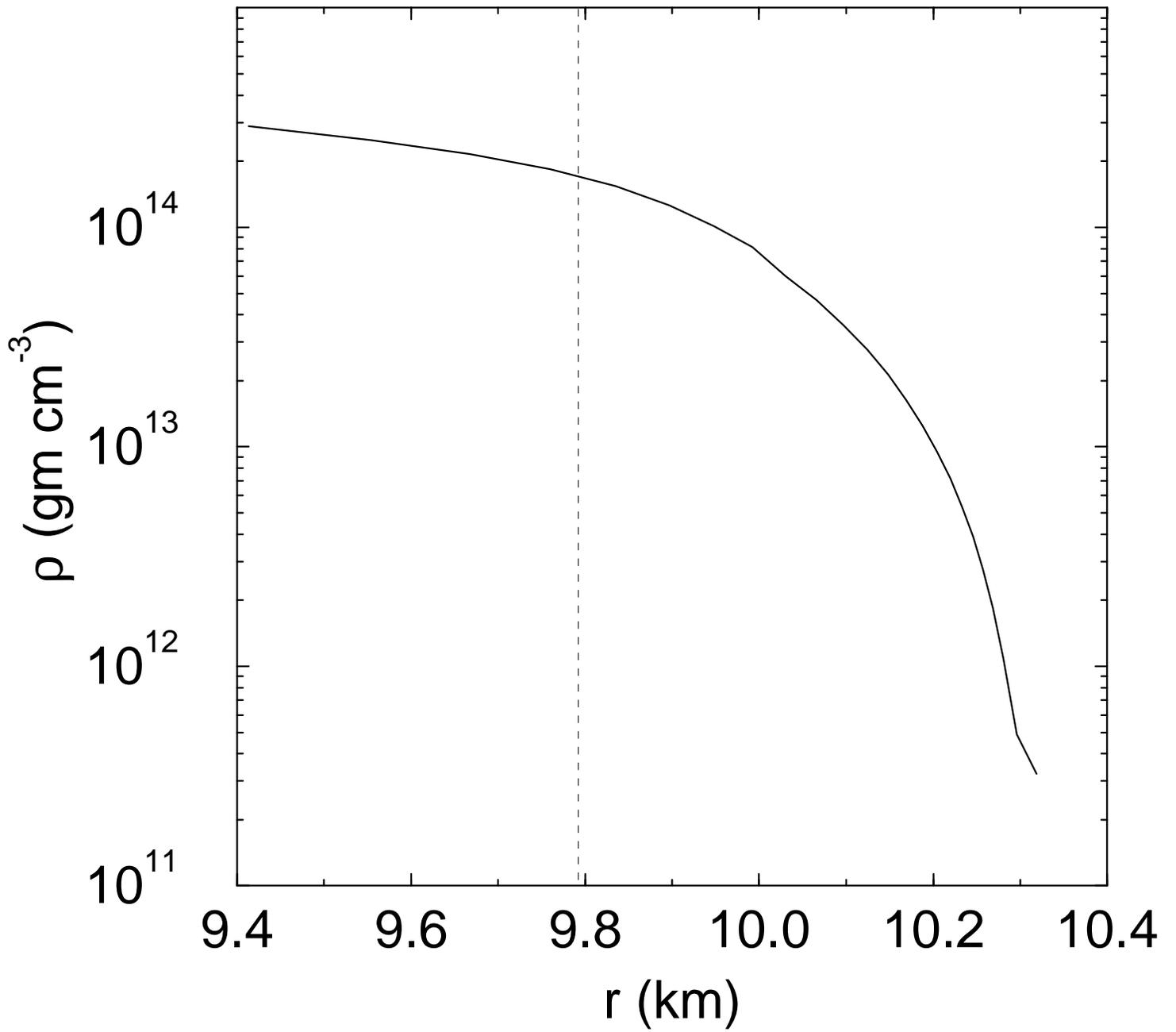

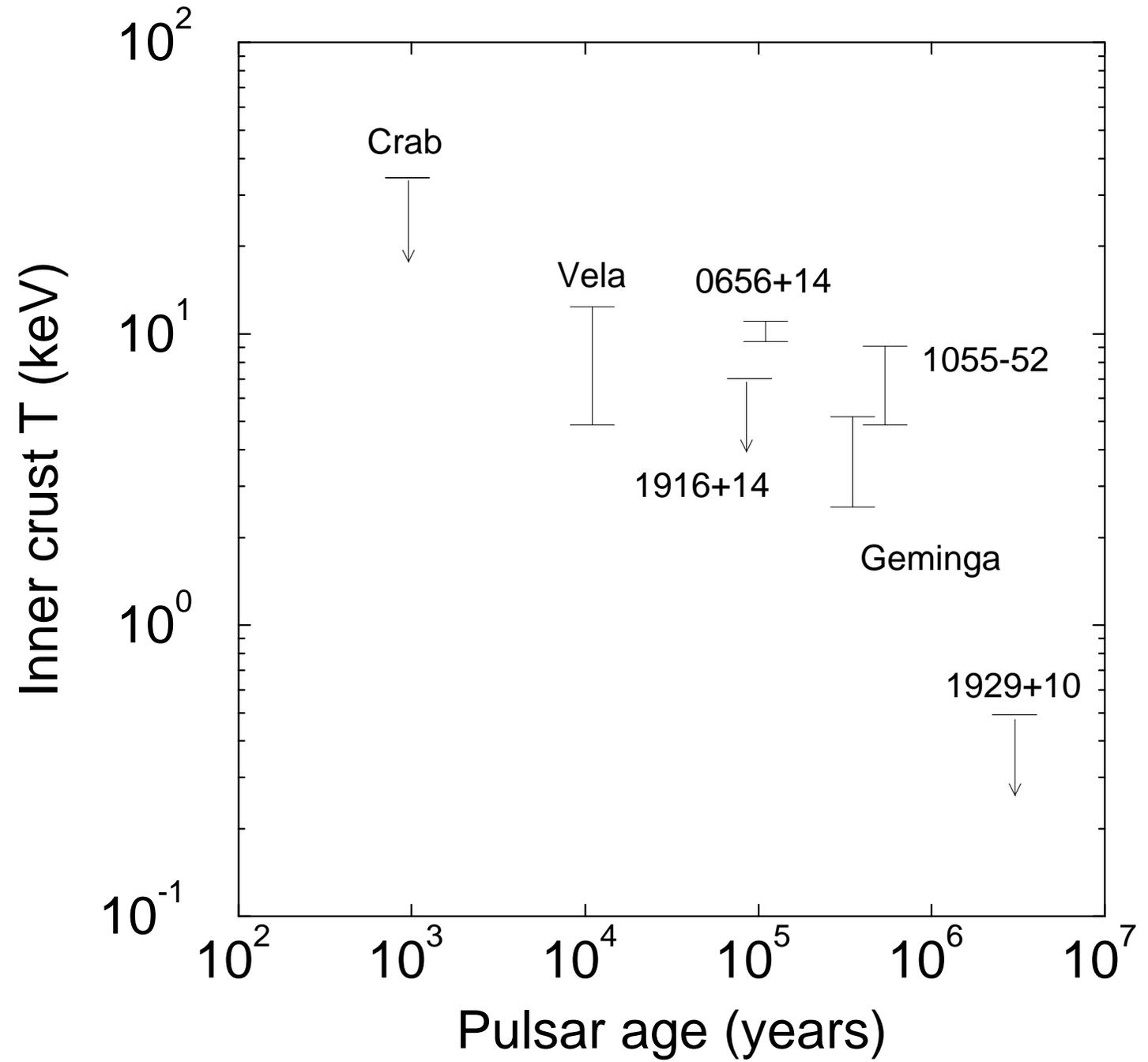

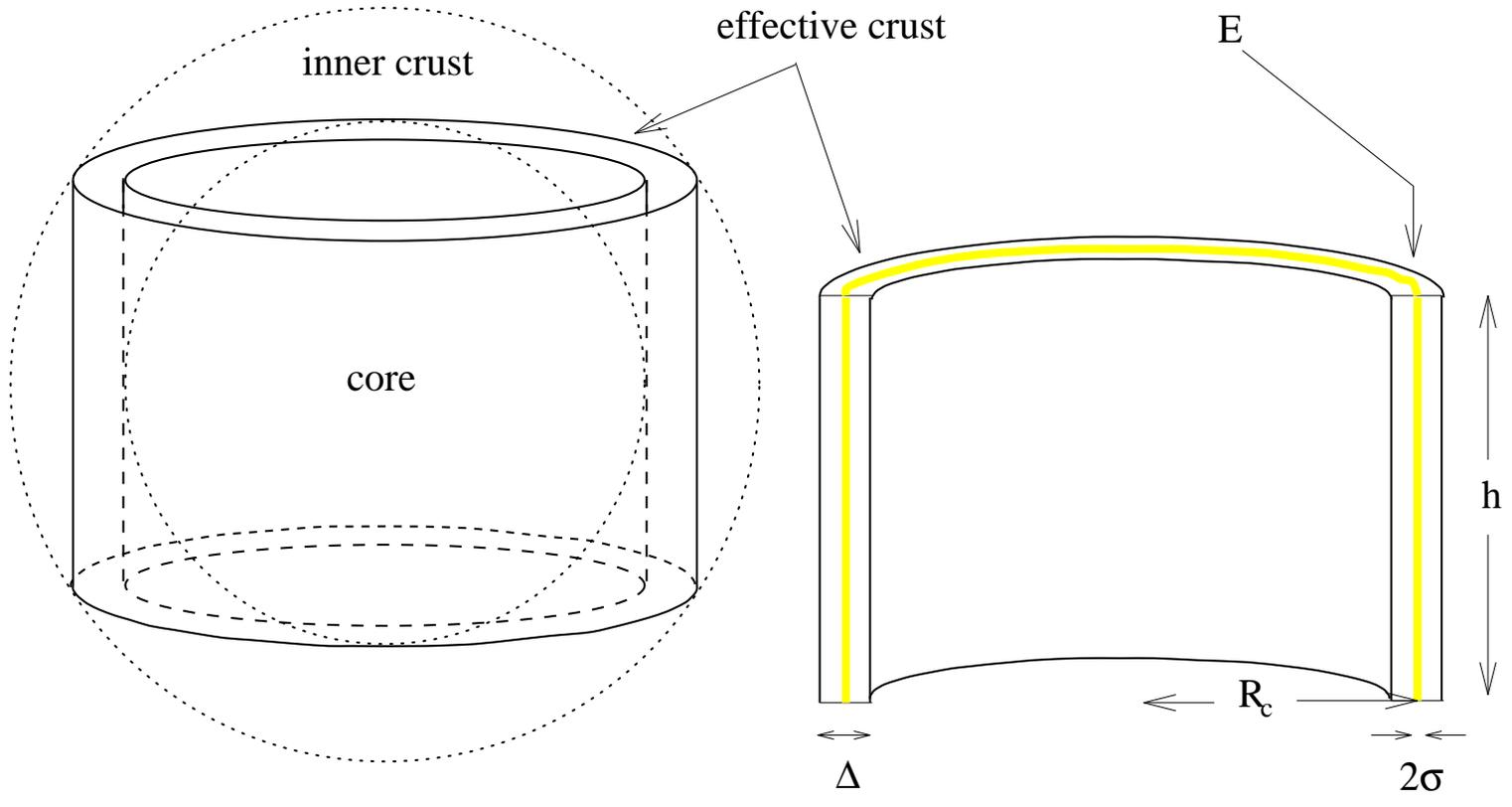

# Crab-like Simulation

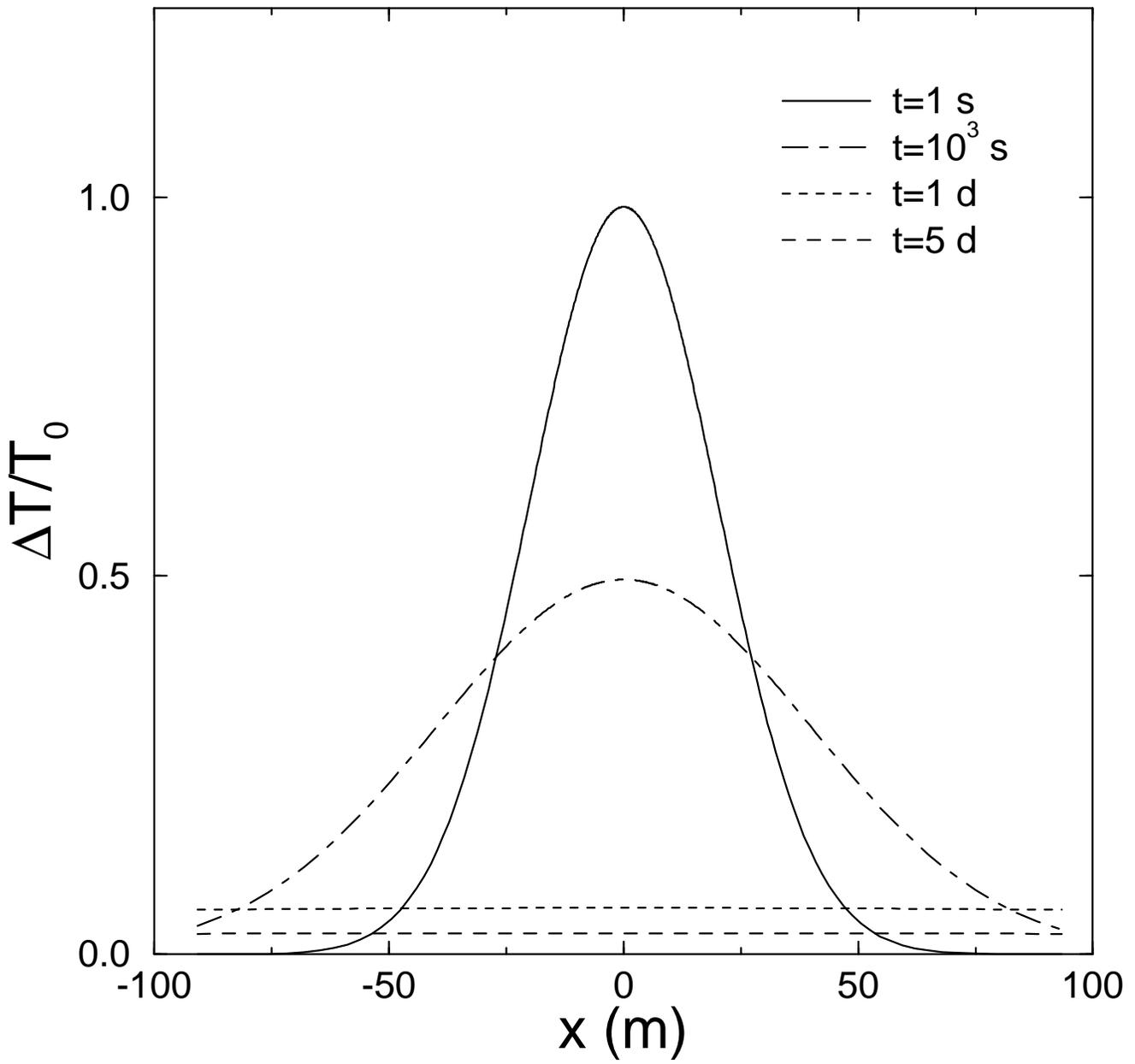

Figure 4

## Crab-like Simulation

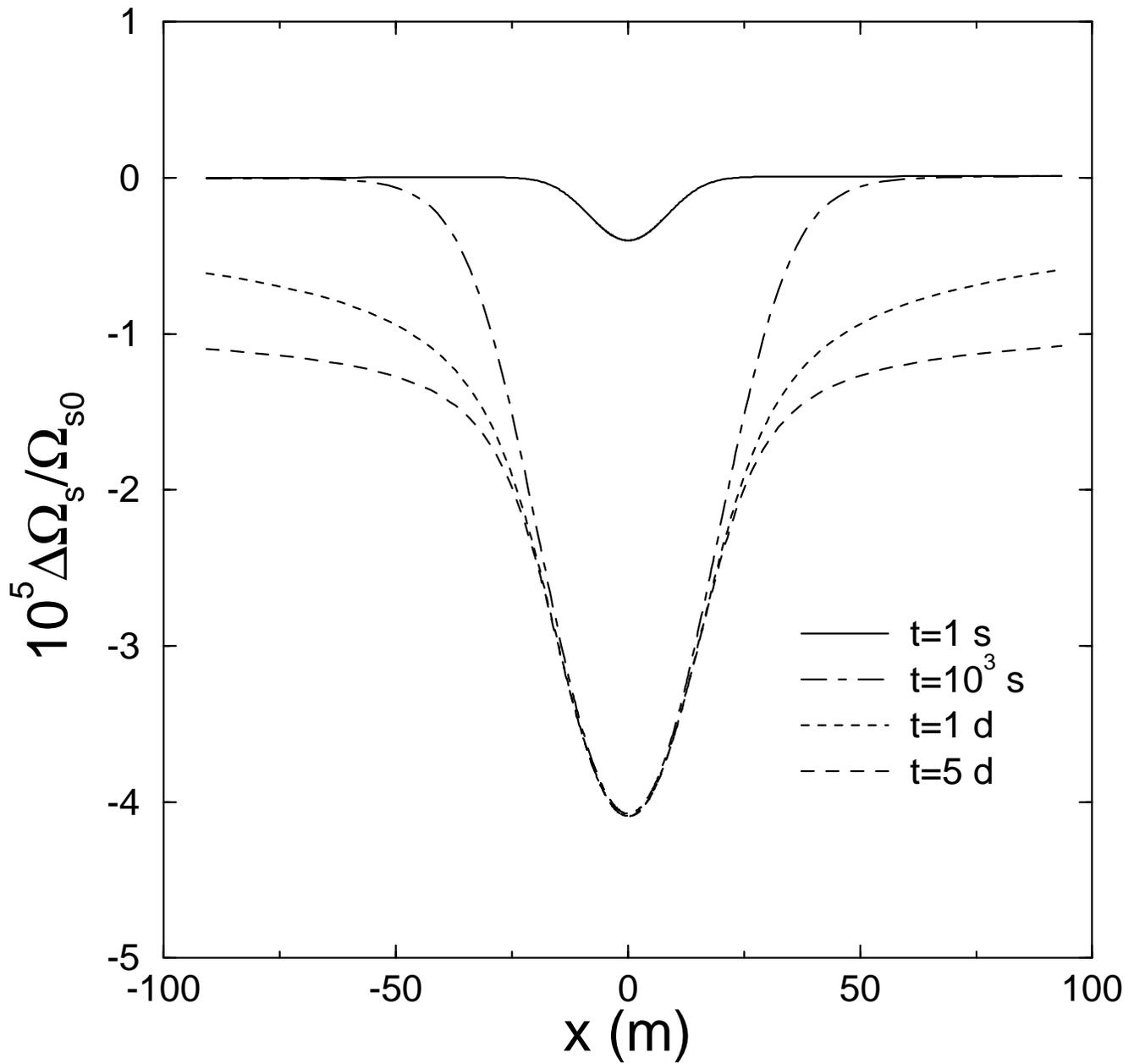

# Crab-like Simulation

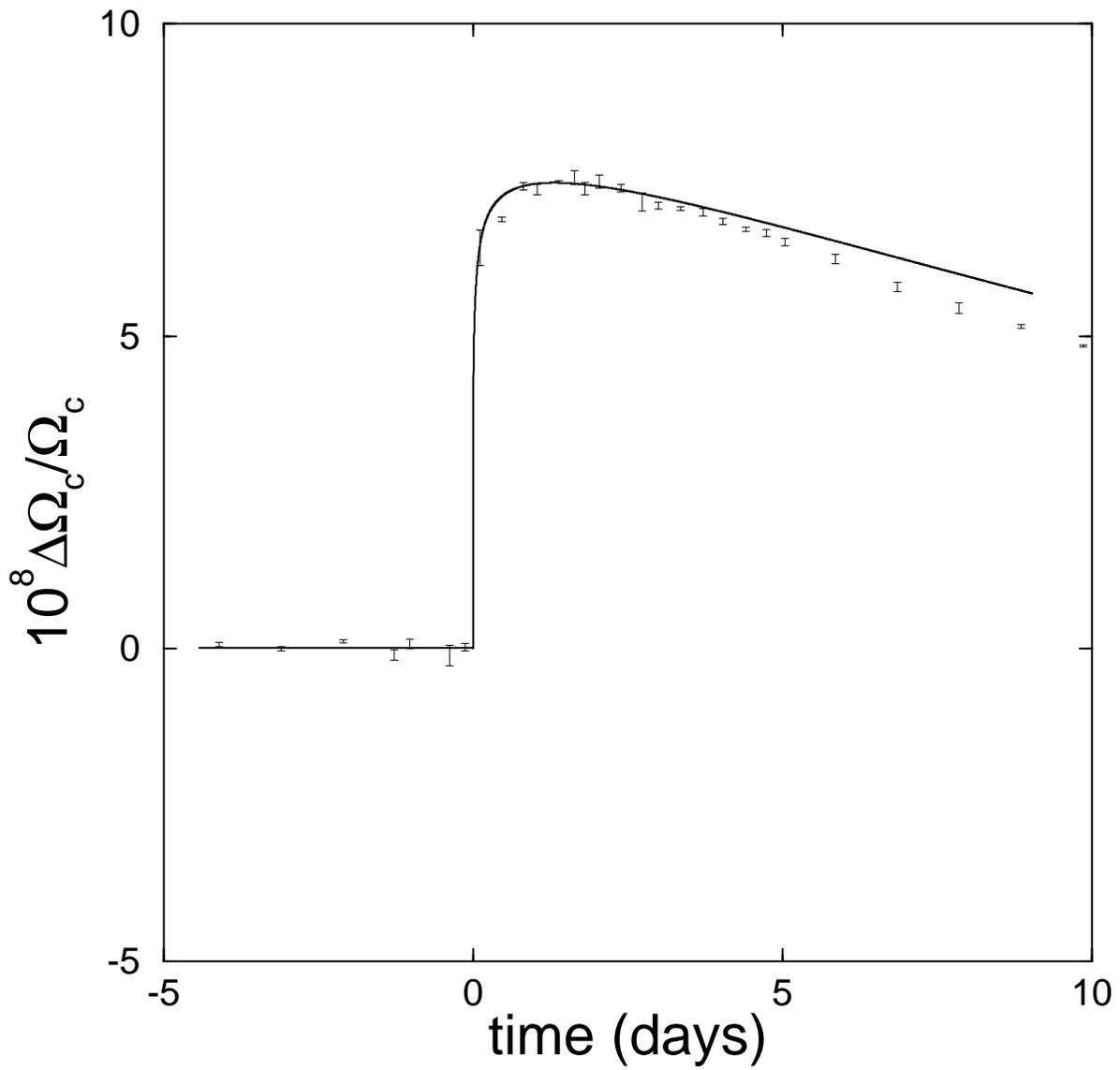

Figure 6

# Crab-like Simulation

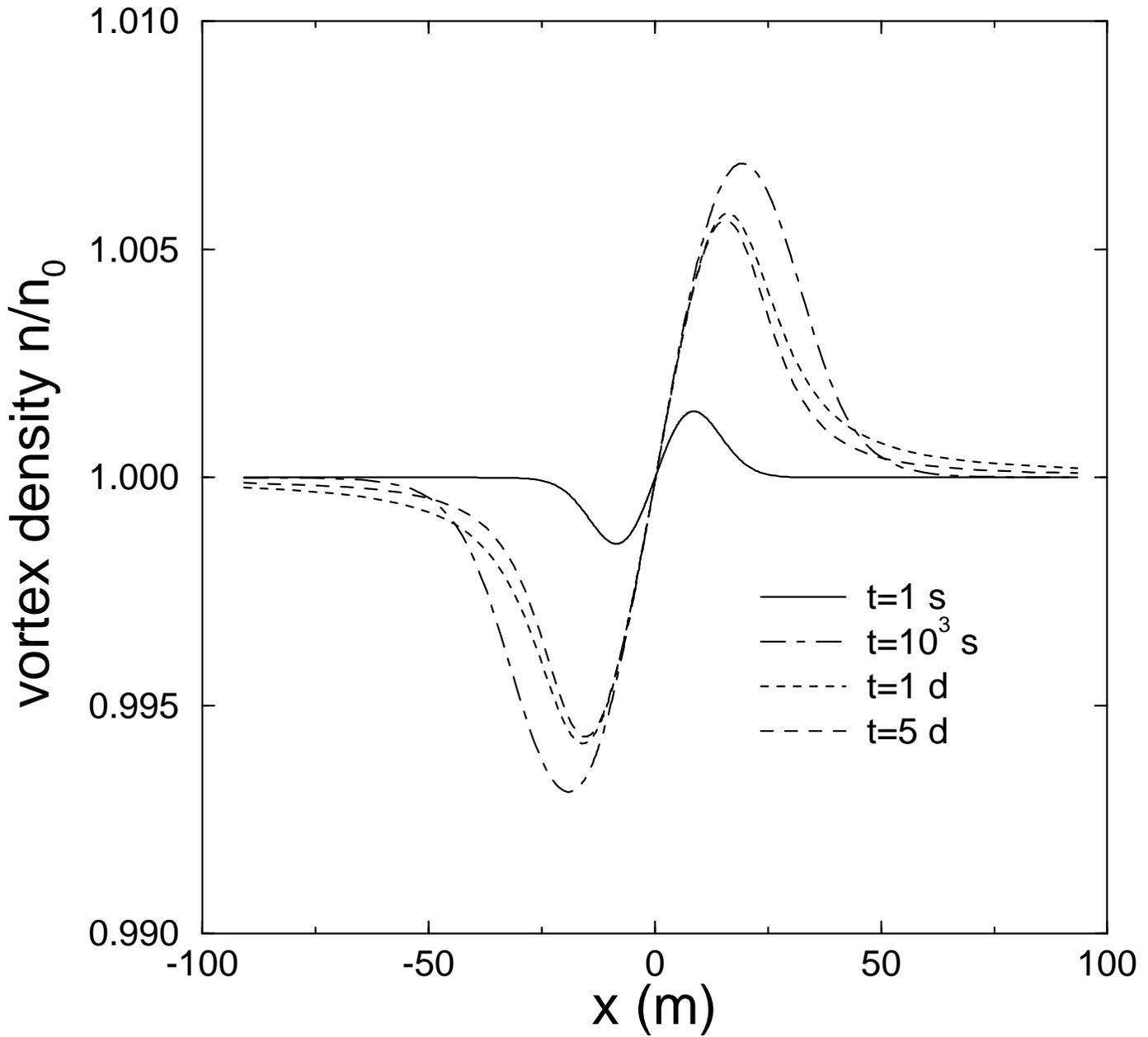

# Vela-like Simulation

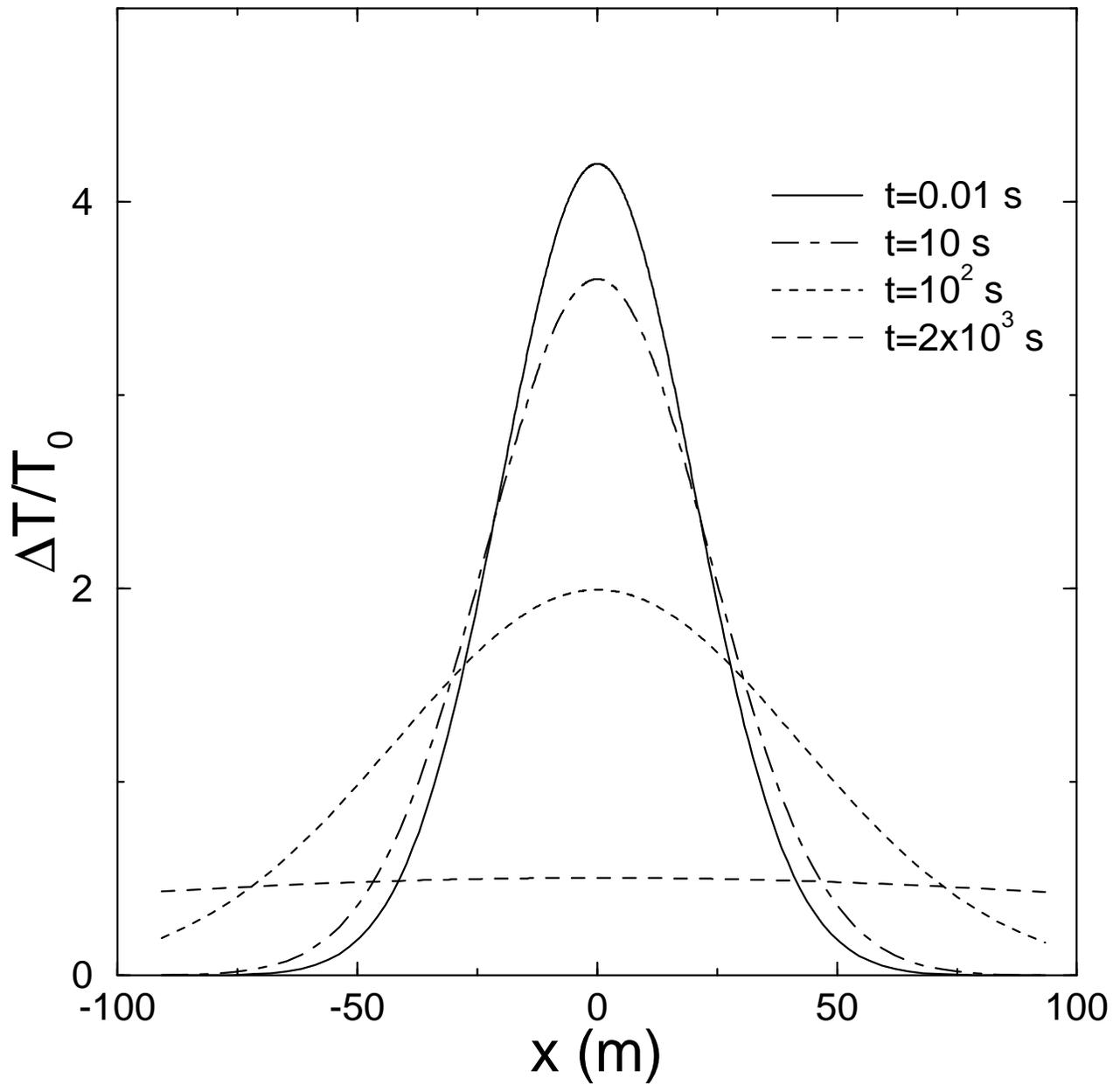

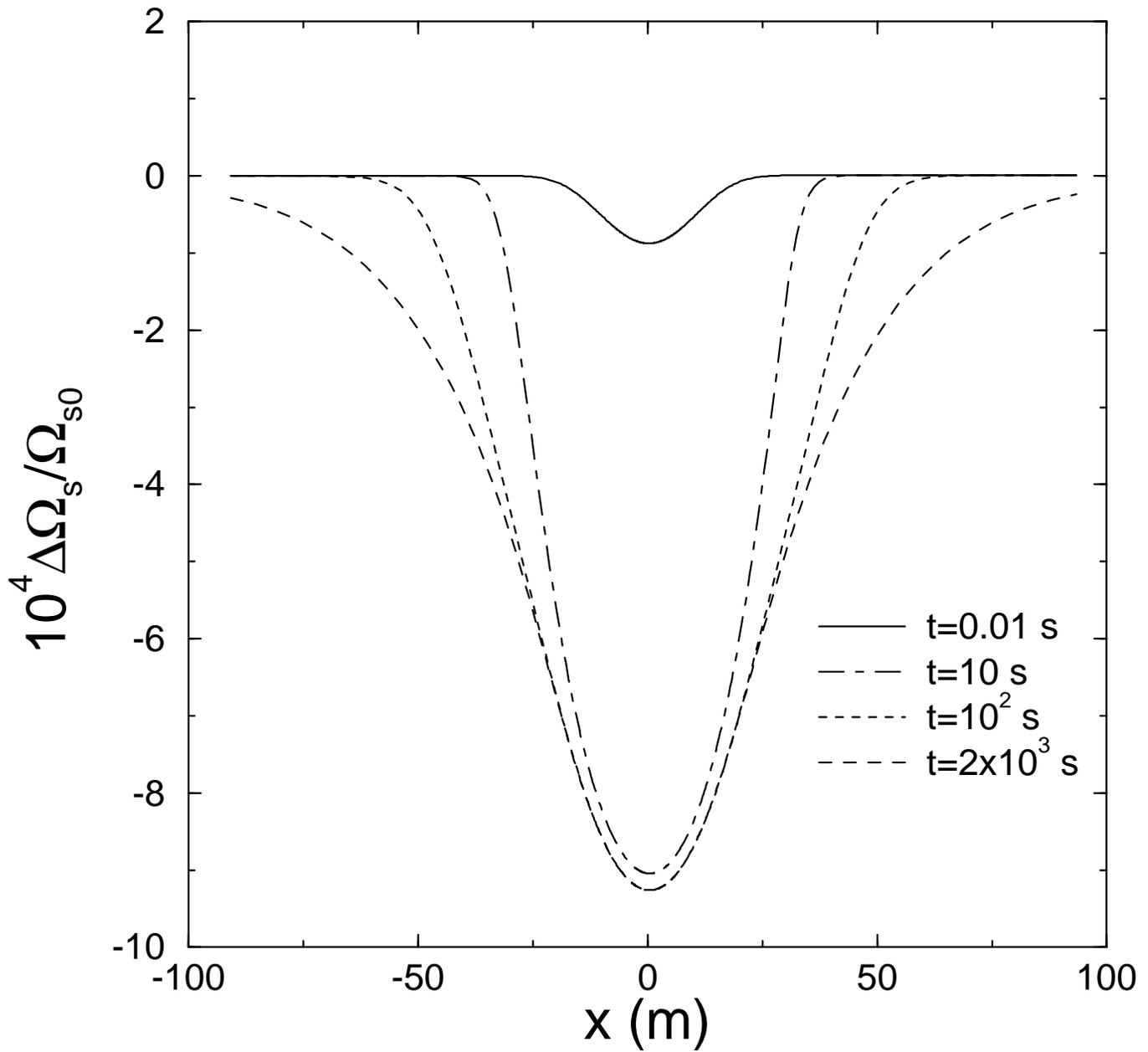

Vela-like Simulation

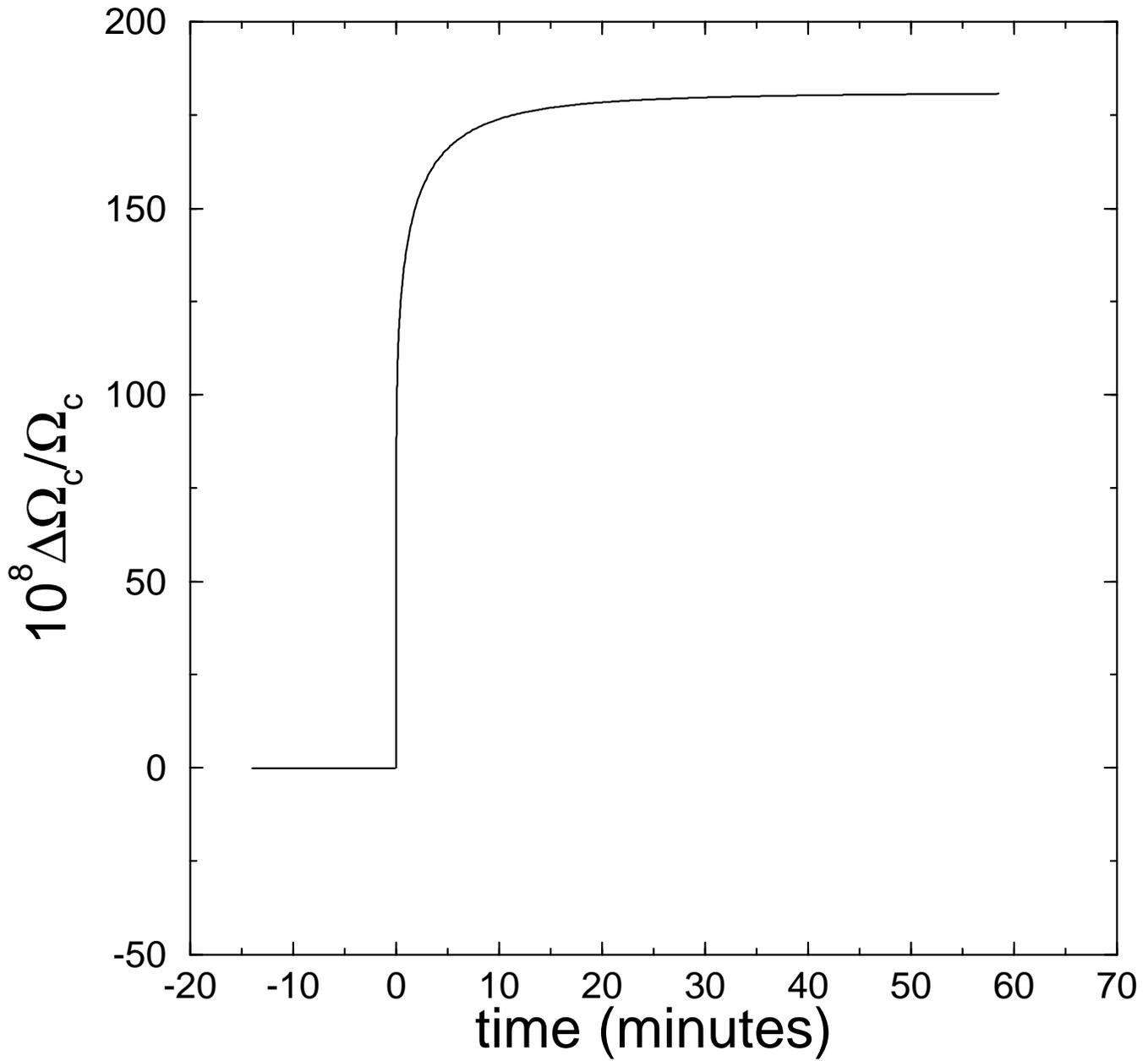

# Vela-like Simulation

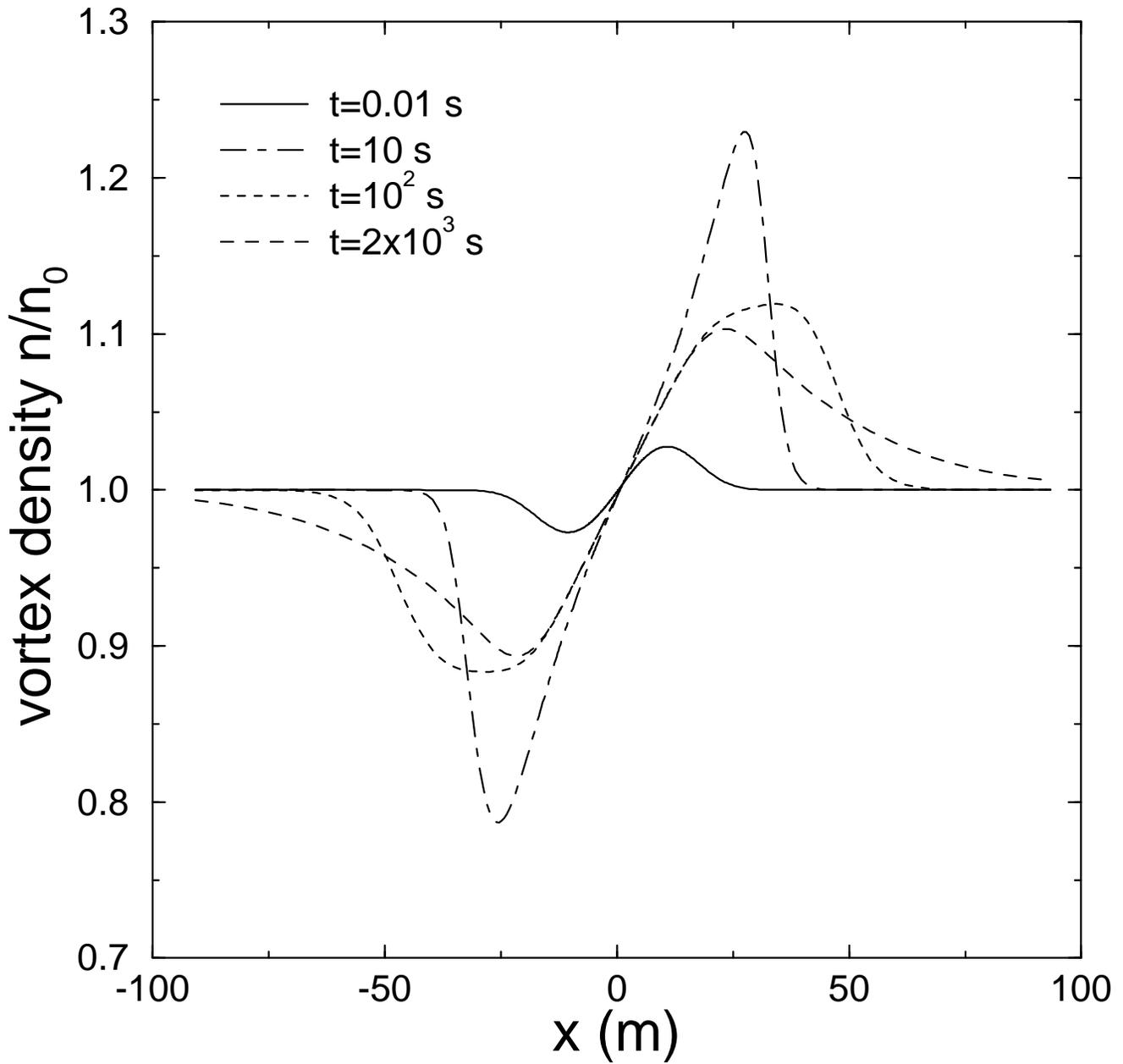

# Vela-like Simulation

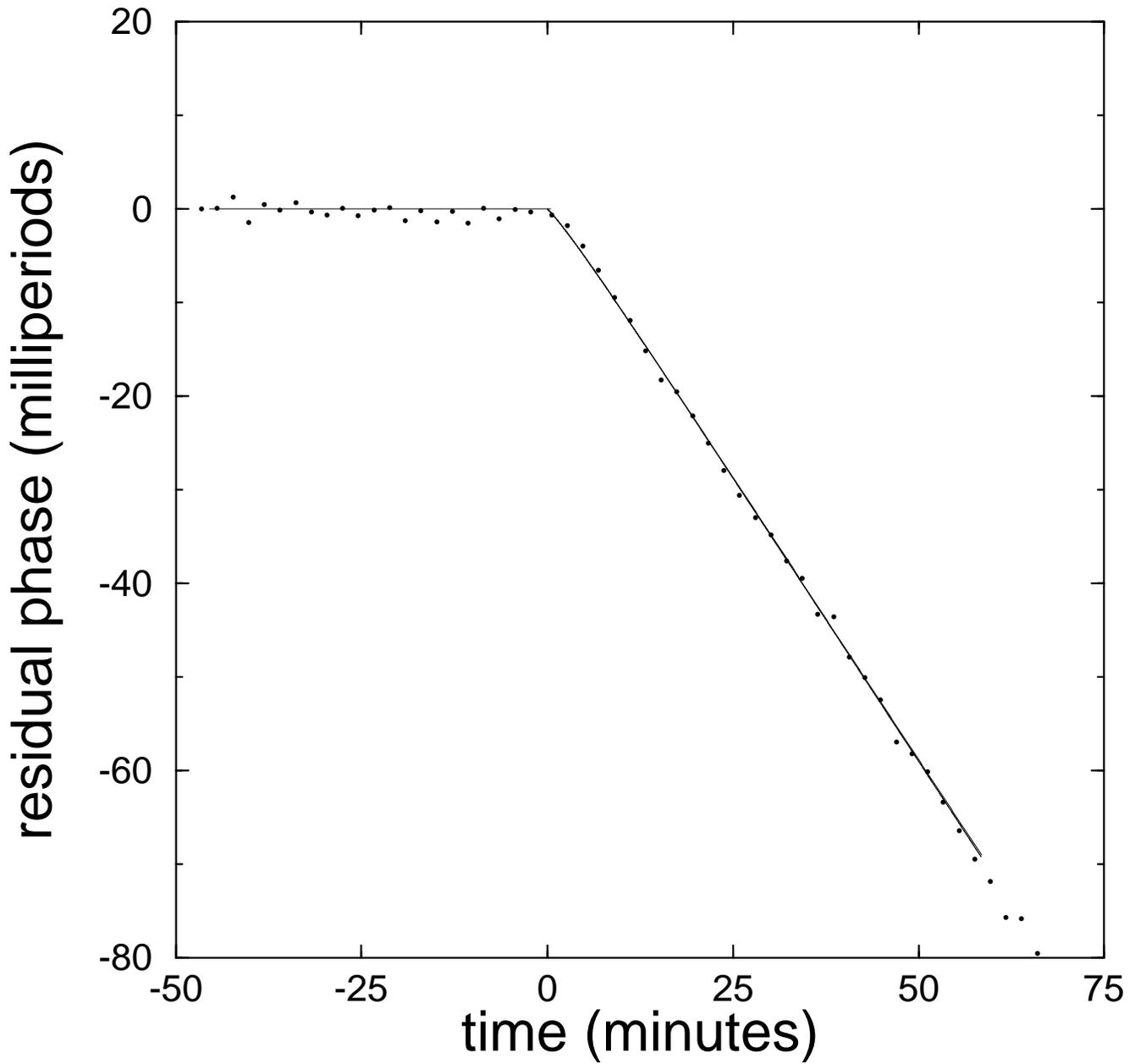

Figure 12

# Vela-like Simulation

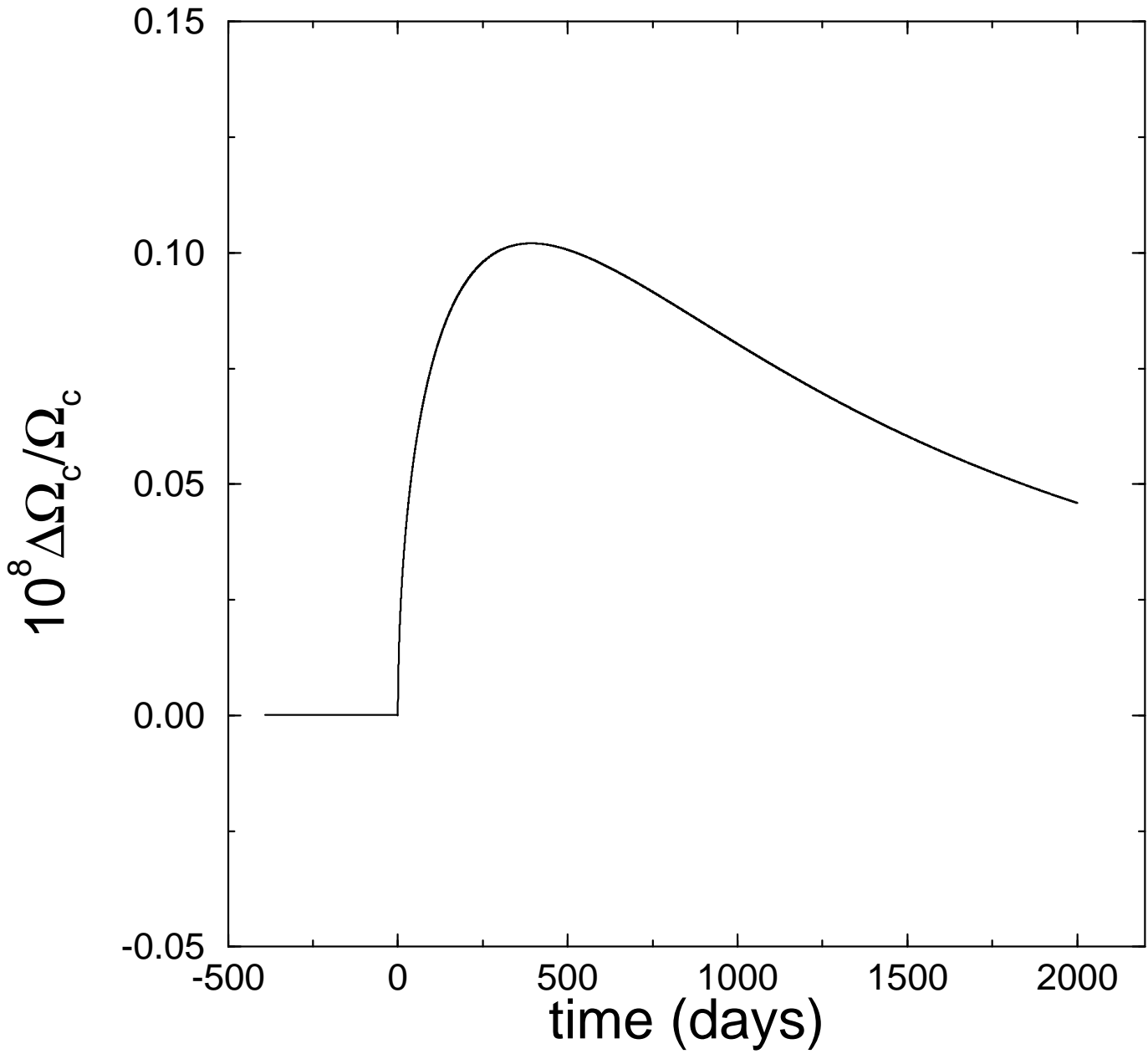

# Older Star

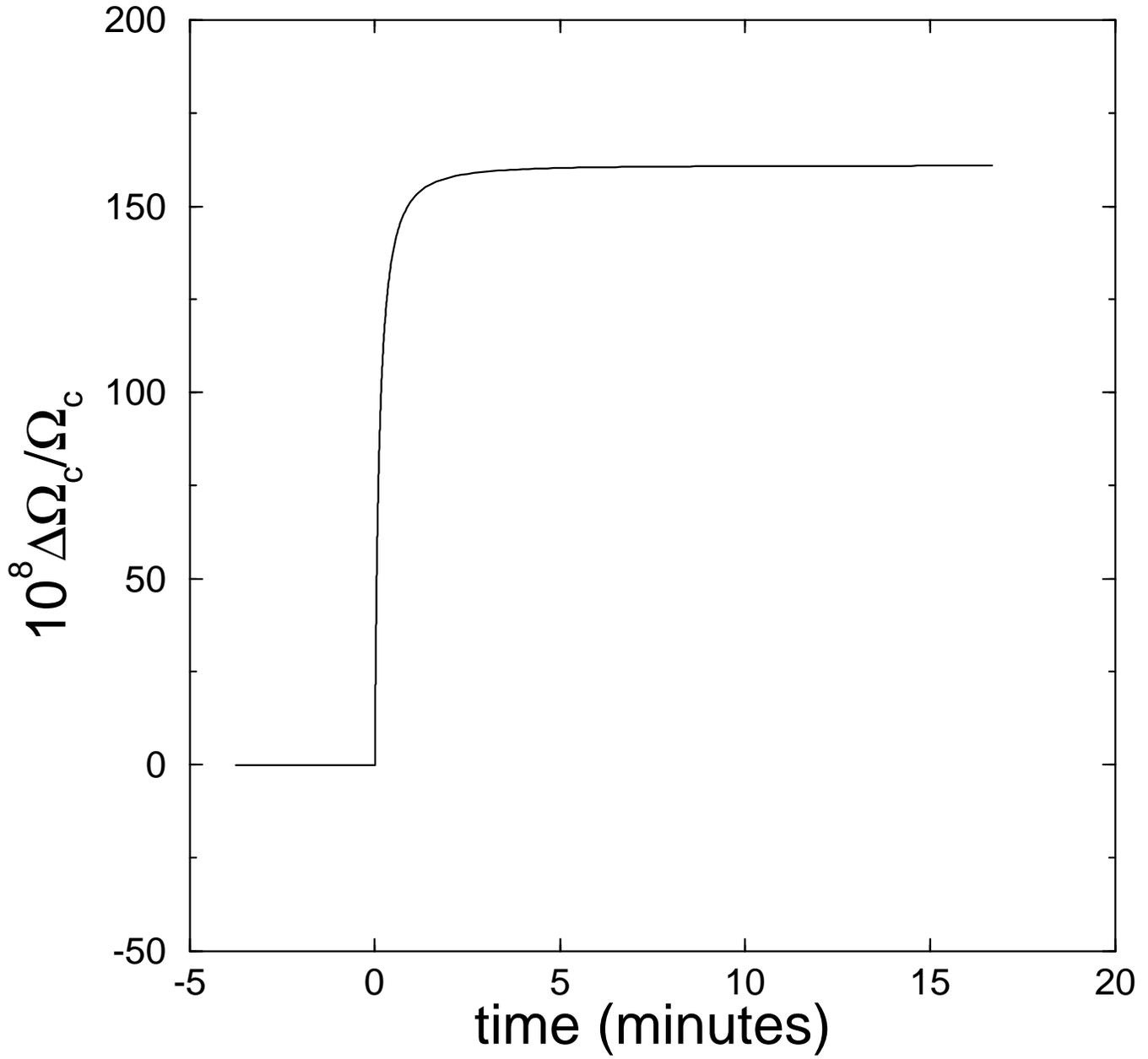

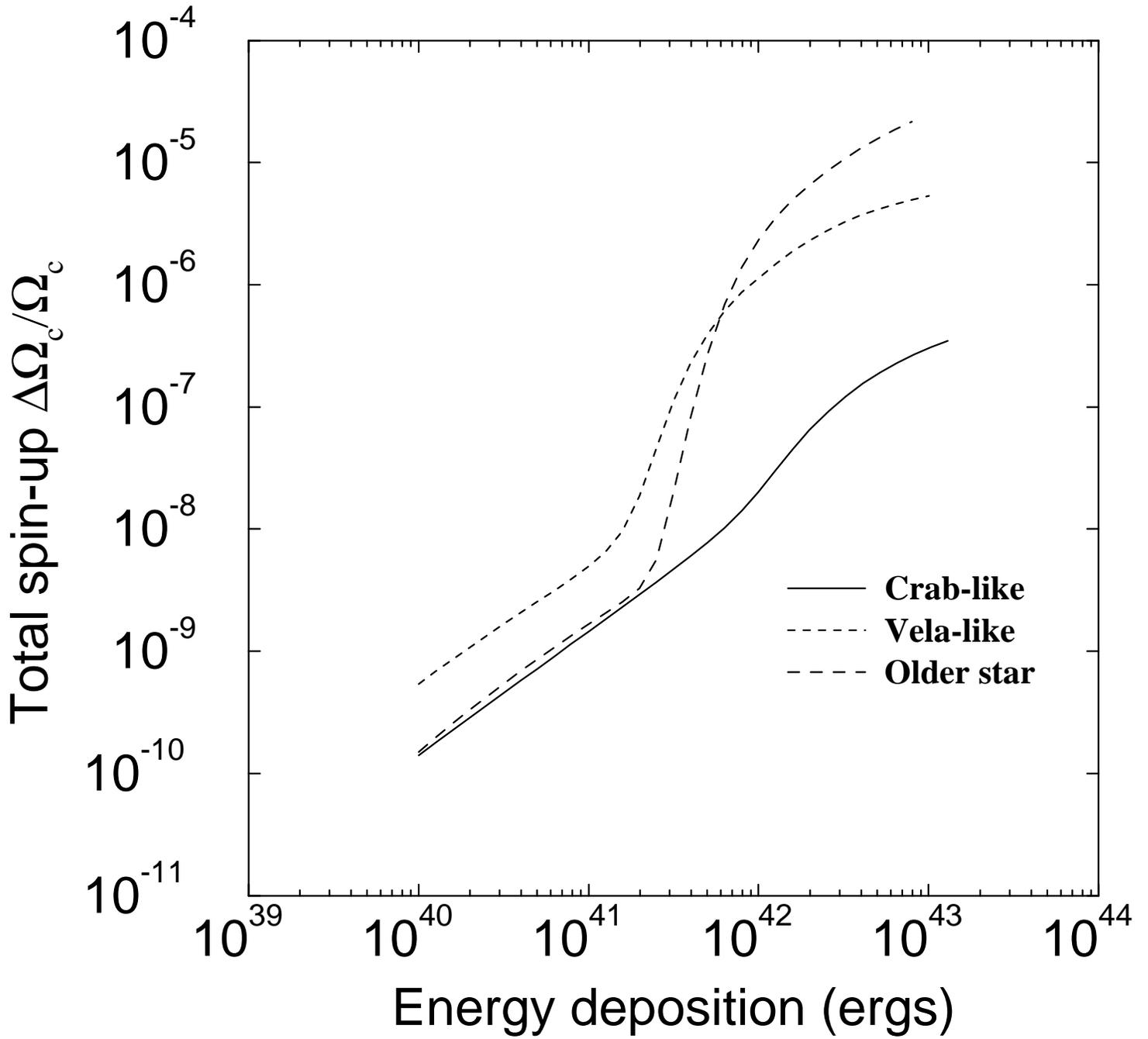

Figure 15

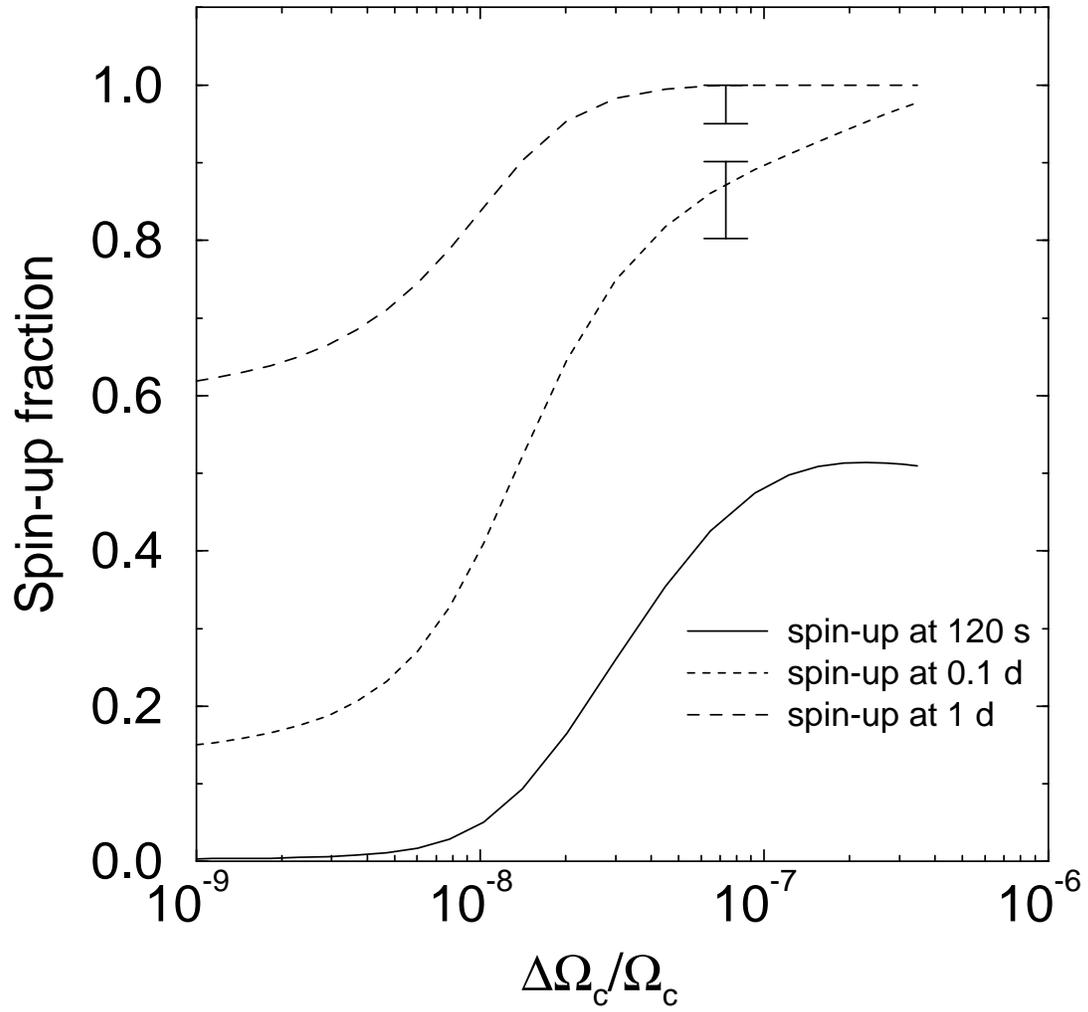

Figure 16

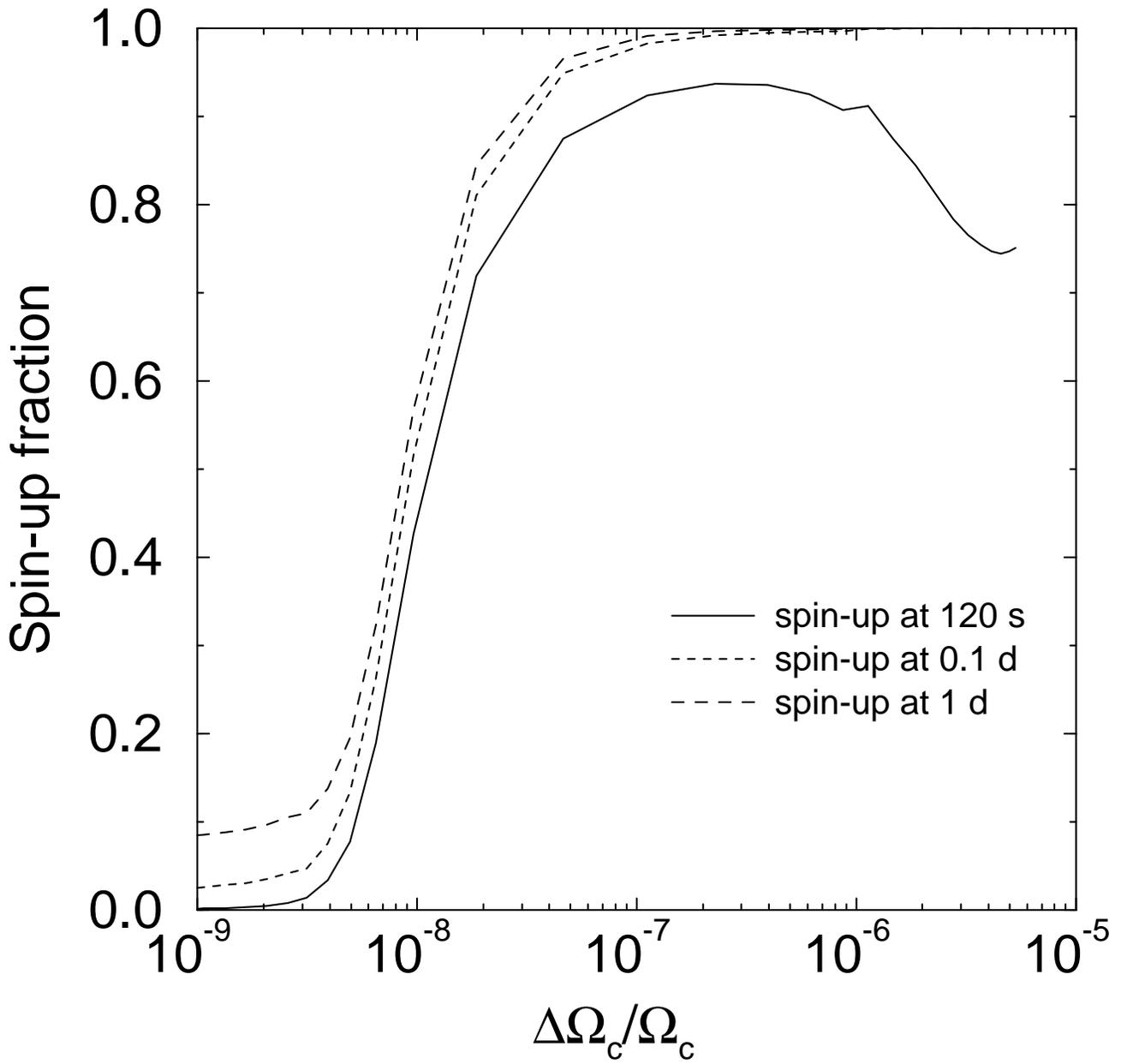

Figure 17

TABLE 1
MODEL SIMULATION PARAMETERS

| Model | $\Omega_c$ rad s$^{-1}$ | $|\dot{\Omega}_c|$ rad s$^{-2}$ | $T_0$ K (keV) | $C_{v0}$ erg cm$^{-3}$ K$^{-1}$ | $\kappa_0$ erg cm$^{-1}$ s$^{-1}$ K$^{-1}$ |
|---|---|---|---|---|---|
| Crab-like | 189 | $2.4 \times 10^{-9}$ | $2.1 \times 10^8$ (18) | $9.5 \times 10^{17}$ | $5.7 \times 10^{21}$ |
| Vela-like | 70.4 | $1.0 \times 10^{-10}$ | $6.1 \times 10^7$ (5.3) | $2.8 \times 10^{17}$ | $1.9 \times 10^{22}$ |
| Older star[a] | $10.5^b$ | $3.2 \times 10^{-13}$ | $6.1 \times 10^7$ (5.3) | $2.8 \times 10^{17}$ | $1.9 \times 10^{22}$ |

[a]Spin-down age $5 \times 10^5$ years.

[b]Scaled from Vela's spin rate with an $n = 3$ breaking index model.

TABLE 2
MODEL SIMULATION PARAMETERS

| $R_c$ (km) | $I$ (g cm$^2$) | $\bar{\rho}$ ( g cm$^{-3}$) | $\beta$ (rad s$^{-1}$ K) | $\Delta$ (m) | $h$ (km) | $\sigma$ (m) | $v_0$ (cm s$^{-1}$) |
|---|---|---|---|---|---|---|---|
| 9.9 | $8.6 \times 10^{44}$ | $1.5 \times 10^{14}$ | $2.0 \times 10^8$ | 200 | 2.8 | 20 | $10^6$ |

# REFERENCES

Abney, M., Epstein, R. I., & Olinto, A. 1995, *in preparation.*

Ainsworth, T., Pines, D., & Wambach, J. 1989, Phys. Lett. B, 222, 173.

Alpar, M. A. 1977, ApJ, 213, 527.

Alpar, M. A., Anderson, P. W., Pines, D. & Shaham, J. 1984, ApJ, 276, 325.

Alpar, M. A., Brinkmann, W., Kiziloğlu, Ü., & Ögelman, H. 1987, A&A, 177, 101.

Alpar, M. A., Chau, H. F., Cheng, K. S., & Pines, D. 1994, ApJ, 427, L29.

Alpar, M. A., Cheng, K. S., & Pines, D. 1989, ApJ, 346, 823.

Alpar, M. A., Langer, S. A., & Sauls, J. A. 1984, ApJ, 282, 533.

Anderson, P. W., Alpar, M. A., Pines, D., & Shaham, J. 1982, Phil. Mag. A., 45, 227.

Anderson, P. W. & Itoh, N. 1975, Nature, 256, 25.

Baym, G. & Pines, D. 1971, Ann. Phys., 66, 816.

Bildsten, L. & Epstein, R. I. 1989, ApJ, 342, 951.

Boynton, P. E. 1981, in *IAU Symposium 95, Pulsars*, ed. R. Wielebinski & W. Sieber (Dordrecht: Reidel), p. 279.

Boynton, P. E. & Deeter, J. E. 1979, in *Compact Galactic X-Ray Sources*, ed. F. K. Lamb & D. Pines (Urbana: University of Illinois), p. 168.

Boynton, P. E., Deeter, J. E., Lamb, F. K., Zylstra G., Pravdo, S. H., White, N. E., Wood. K. S., & Yentis, D. J. 1984, ApJ, 283, L53.

Chau, H. F. & Cheng, K. S. 1993a, in *Isolated Pulsars*, Eds. K. A. Van Riper, R. I. Epstein, & C. Ho (Cambridge: Cambridge), p35.

——1993b, Phys. Rev. B, 47, 2707.

Cheng, K. S., Chau, W. Y., Zhang, J. L., & Chau, H. F. 1992, ApJ, 396, 135.

Ding, K. Y., Cheng, K. S., & Chau, H. F. 1993, in *Isolated Pulsars*, Eds. K. A. Van Riper, R. I. Epstein, & C. Ho (Cambridge: Cambridge), p35.

Epstein, R. I., & Baym, G. 1988, ApJ, 328, 680.

——1992, ApJ, 387, 276.

Feibelman, P. J. 1971, Phys. Rev. D, 4, 1589.

Friedman, B. & Pandharipande, V. R. 1981, Nucl. Phys. A, 361, 502 (FP).

Greenstein, G. 1975, ApJ, 200, 281.

———1979a, ApJ, 231, 880.

———1979b, Nature, 277, 521.

Harding, D., Guyer, R. A., & Greenstein, G. 1978, ApJ, 222, 991.

Gudmundsson, E., Pethick, C., & Epstein, R. I. 1982, ApJ, 259, L19.

Jones, P. B. 1987, MNRAS, 228, 513.

———1990a, MNRAS, 243, 257.

———1990b, MNRAS, 244, 675.

———1990c, MNRAS, 246, 315.

———1991, ApJ, 373, 208.

———1992, MNRAS, 257, 501.

Link, B. & Epstein, R. I. 1991, ApJ, 373, 592 (LE).

Link, B., Epstein, R. I., & Baym, G. 1993, ApJ, 403, 285 (LEB).

Link, B., Epstein, R. I., & Van Riper, K. A. 1992, Nature, 359, 616.

Lorenz, C. P., Ravenhall, D. G., & Pethick, C. J. 1993, Phys. Rev. Lett., 70, 379.

Lyne, A. G., Smith, F. G., & Pritchard, R. S. 1992, Nature, 359, 706.

McCulloch, P. M., Hamilton, P. A., McConnell, D., & King, E. A. 1990, Nature, 346, 822.

Miller, M. C. 1992, MNRAS, 255, 129.

———1993, in *Isolated Pulsars*, Eds. K. A. Van Riper, R. I. Epstein, & C. Ho (Cambridge: Cambridge), p153.

Pethick, C. J., Ravenhall, D. G., Lorenz, C. P. 1995, Nuc.Phys.A, 584, 675.

Pines, D. & Shaham, J. 1972, Nat. Phys. Sci, 235, 43.

Pines, D., Shaham, J., & Ruderman, M. 1972, Nat. Phys. Sci, 237. 83.

Romani, R. W. 1987, ApJ, 313, 718.

Ruderman, M. 1969, Nature, 223, 597.

—–1976, ApJ, 203, 213.

Shibanov, Yu. A., Zavlin V. E., Pavlov, G. G., Ventura, J., & Potekhin, A. Yu. 1993, in *Isolated Pulsars*, Eds. K. A. Van Riper, R. I. Epstein, & C. Ho (Cambridge: Cambridge), p174.

Shibazaki, N. & Lamb, F. K. 1989, ApJ, 346, 808.

—–, in preparation.

Shibazaki N. & Mochizuki Y. 1994, ApJ, 438, 288.

Srinivasan, G., Bhattacharya, D., Muslimov, A., Asygan, A., 1990, Current Sci., 59, 31.

Wolszczan, A. 1991, Nature, 350, 688.

Umeda, H., Shibazaki, N., Nomoto, K., & Tsuruta, S. 1993, ApJ, 408, 186.

Urpin, V. A. & Van Riper, K. A. 1993, ApJ, 411, L87.

Van Riper, K. A. 1991, ApJ, 372, 251.

Van Riper, K. A., Link, B., & Epstein, R. I. 1995, ApJ, 448, 294.

Van Riper, K. A., Epstein, R. I., & Miller, G. S. 1991, ApJ, 381, L47.

Wiringa, R. B., Fiks, V., & Fabrocini, A. 1988, Phys. Rev. C, 38, 1010.